\documentclass[final,3p,times]{elsarticle}
\usepackage{hyperref}
\usepackage{amsmath,amsfonts,amssymb,amsthm}
\usepackage{tikz}
  \usetikzlibrary{arrows,automata,shapes}
  \usetikzlibrary{backgrounds}
  \usetikzlibrary{decorations.pathmorphing}
  \tikzset{snake it/.style={-stealth,decoration={snake,amplitude=.4mm,segment length=2mm,post length=1.9mm},decorate}}
\usepackage{microtype}
\usepackage{enumerate}
\usepackage{mathtools}
\usepackage{relsize}
\usepackage{mathrsfs}
\usepackage{wrapfig}
\usepackage{xspace}

\allowdisplaybreaks

\newcommand{\eps}{\varepsilon}
\DeclareMathOperator{\enc}{\mathrm enc}
\DeclareMathOperator{\enctwo}{\mathrm enc}
\DeclareMathOperator{\sub}{\mathrm sub}
\DeclareMathOperator{\alp}{\rm alph}
\DeclareMathOperator{\A}{\mathcal A}
\DeclareMathOperator{\B}{\mathcal B}
\DeclareMathOperator{\D}{\mathcal D}
\DeclareMathOperator{\C}{\mathcal C}
\DeclareMathOperator{\M}{\mathcal M}
\DeclareMathOperator{\reach}{\mathrm reach}
\DeclareMathOperator{\R}{\mathcal{R}}
\DeclareMathOperator{\J}{\mathcal{J}}
\DeclareMathOperator{\depth}{\mathrm depth}

\newcommand{\tuple}[1]{\langle{#1}\rangle}

\newcommand{\tmInputAlphabet}{I}
\newcommand{\blank}{\llcorner\!\lrcorner}

\newcommand{\complclass}[1]{{\sc #1}\xspace}
\newcommand{\NL}{\complclass{NL}}

\newcommand{\coNP}{\complclass{coNP}}
\newcommand{\PSpace}{\complclass{PSpace}}
\newcommand{\complete}{comp.}
\newcommand{\columnspace}{~~~~~~}
\newcommand{\Deltaplus}{\Delta_{\#\$}}

\theoremstyle{plain}
\newtheorem{theorem}{Theorem}
\newtheorem{lemma}[theorem]{Lemma}
\newtheorem{corollary}[theorem]{Corollary}

\newtheorem{claim}[theorem]{Claim}

\newdefinition{remark}[theorem]{Remark}
\newdefinition{example}[theorem]{Example}
\newdefinition{definition}[theorem]{Definition}
\newproof{pf}{Proof}

\journal{}

\begin{document}
\begin{frontmatter}

\title{Complexity of Universality and Related Problems for Partially Ordered NFAs}

\author[tud]{Markus Kr\"otzsch\fnref{sup}}
\ead{markus.kroetzsch@tu-dresden.de}

\author[tud,mu]{Tom\'{a}\v{s} Masopust\corref{cor1}\fnref{sup}}
\ead{masopust@math.cas.cz}

\author[inria]{Micha\"el Thomazo}
\ead{michael.thomazo@inria.fr}

\fntext[tm]{This work was supported by the German Research Foundation (DFG) in Emmy Noether grant KR~4381/1-1 (DIAMOND).}

\address[tud]{Institute of Theoretical Computer Science and Center of Advancing Electronics Dresden (cfaed), TU Dresden, Germany}

\address[mu]{Institute of Mathematics, Czech Academy of Sciences, {\v Z}i{\v z}kova 22, 616 62 Brno, Czechia}

\address[inria]{Inria, France}

\cortext[cor1]{Corresponding author}

\begin{abstract}
  Partially ordered nondeterministic finite automata (poNFAs) are NFAs whose transition relation induces a partial order on states, that is, for which cycles occur only in the form of self-loops on a single state. A poNFA is universal if it accepts all words over its input alphabet. Deciding universality is \PSpace-complete for poNFAs, and we show that this remains true even when restricting to a fixed alphabet. This is nontrivial since standard encodings of alphabet symbols in, e.g., binary can turn self-loops into longer cycles. A lower \coNP-complete complexity bound can be obtained if we require that all self-loops in the poNFA are deterministic, in the sense that the symbol read in the loop cannot occur in any other transition from that state. We find that such restricted poNFAs (rpoNFAs) characterize the class of $\R$-trivial languages, and we establish the complexity of deciding if the language of an NFA is $\R$-trivial. Nevertheless, the limitation to fixed alphabets turns out to be essential even in the restricted case: deciding universality of rpoNFAs with unbounded alphabets is \PSpace-complete. Based on a close relation between universality and the problems of inclusion and equivalence, we also obtain the complexity results for these two problems. Finally, we show that the languages of rpoNFAs are definable by deterministic (one-unambiguous) regular expressions, which makes them interesting in schema languages for XML data.
\end{abstract}


\end{frontmatter}

\section{Introduction}
  The universality problem asks if a given automaton (or grammar) accepts (or generates) all possible words over its alphabet. In typical cases, deciding universality is more difficult than deciding the word problem. For example, universality is undecidable for context-free grammars~\cite{BHPS61} and \PSpace-complete for nondeterministic finite automata (NFAs)~\cite{MeyerS72}. The study of universality (and its complement, emptiness) has a long tradition in formal languages, with many applications across computer science, e.g., in the context of formal knowledge representation and database theory \cite{BarceloLR:jacm14,CalvaneseGLV03:rpqreasoning,SMKR:elcq14}. Recent studies investigate the problem for specific types of automata or grammars, e.g., for prefixes or factors of regular languages~\cite{RampersadSX12}.

  In this paper, we are interested in the universality problem for \emph{partially ordered NFAs} (poNFAs) and special cases thereof. An NFA is partially ordered if its transition relation induces a partial order on states: the only cycles allowed are self-loops on a single state. Partially ordered NFAs define a natural class of languages that has been shown to coincide with level~$\frac{3}{2}$ of the Straubing-Th\'erien hierarchy~\cite{SchwentickTV01} and with Alphabetical Pattern Constraint (APC) languages, a subclass of regular languages effectively closed under permutation rewriting~\cite{BMT2001}. Deciding whether an automaton recognizes an APC language (and hence whether it can be recognized by a poNFA) is \PSpace-complete for NFAs and \NL-complete for DFAs~\cite{BMT2001}.
  
  Restricting to partially ordered deterministic finite automata (poDFAs), we can capture further classes of interest: two-way poDFAs characterize languages whose syntactic monoid belongs to the variety {\bf DA}~\cite{SchwentickTV01}, introduced by Sch\"utzenberger~\cite{Sch76}; poDFAs characterize $\R$-trivial languages~\cite{BrzozowskiF80}; and confluent poDFAs characterize level~1 of the Straubing-Th\'erien hierarchy, also known as $\J$-trivial languages or piecewise testable languages~\cite{Simon1972}. Other relevant classes of partially ordered automata include partially ordered B\"uchi automata~\cite{KufleitnerL11} and two-way poDFAs with look-around~\cite{LodayaPS10}.

  The first result on the complexity of universality for poNFAs is readily obtained. It is well known that universality of regular expressions is \PSpace-complete~\cite[Lemma~10.2]{AhoHU74}, and it is easy to verify that the regular expressions used in the proof can be expressed in poNFAs:

  \begin{corollary}[Lemma~10.2 \cite{AhoHU74}]
    The universality problem for poNFAs is \PSpace-complete.
  \end{corollary}
  
  A closer look at the proof reveals that the underlying encoding requires an alphabet of size linear in the input: \PSpace-hardness is not established for alphabets of bounded size. Usually, one could simply encode alphabet symbols $\sigma$ by sequences $\sigma_1\cdots\sigma_n$ of symbols from a smaller alphabet, say $\{0,1\}$. However, doing this requires self-loops $q\stackrel{\sigma}{\to}q$ to be replaced by nontrivial cycles $q\stackrel{\sigma_1}{\to}\cdots\stackrel{\sigma_n}{\to}q$, which are not permitted in poNFAs.

\begin{figure}[h]
    \centering
    \begin{tikzpicture}[baseline,->,>=stealth,shorten >=0pt,node distance=2.5cm,
      every node/.style={fill=white,font=\small},
      state/.style={circle,minimum size=5mm,draw=black,initial text=}]
      \node[state]  (a) {};
      \node[state]  (aa) [right of=a]  {};
      \path
        (a) edge[loop above] node[outer sep=1pt] {$a$} (a)
        (a) edge node {$a$} (aa)
        ;
    \end{tikzpicture}
    \caption{Nondeterministic self-loops -- the forbidden pattern of rpoNFAs}
    \label{fig_bad_pattern}
  \end{figure}
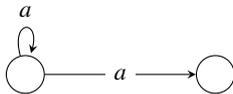

  We settle this open problem by showing that \PSpace-hardness is retained even for binary alphabets. This negative result leads us to ask if there is a natural subclass of poNFAs for which universality does become simpler. We consider \emph{restricted\/} poNFAs (rpoNFAs), which require self-loops to be deterministic in the sense that the automaton contains no transition as in Figure~\ref{fig_bad_pattern}, which we call {\em nondeterministic self-loops\/} in the rest of the paper.
  Large parts of the former hardness proof hinge on transitions of this form, which, speaking intuitively, allow the automaton to navigate to an arbitrary position in the input (using the loop) and, thereafter, continue checking an arbitrary pattern. Indeed, we find that the universality becomes \coNP-complete for rpoNFAs with a fixed alphabet.

  \begin{table}
    \mbox{}\hfill%
      \begin{tabular}{@{}r@{\columnspace}r@{ }l@{\columnspace}r@{ }l@{\columnspace}r@{ }l@{}}
          & \multicolumn{2}{@{}l}{Unary alphabet} 
          & \multicolumn{2}{@{}l}{Fixed alphabet} 
          & \multicolumn{2}{@{}l}{Arbitrary alphabet}\\
        \hline
        DFA     & L-\complete       & \cite{Jones75}
                & \NL-\complete     & \cite{Jones75}
                & \NL-\complete     & \cite{Jones75}\\
        rpoNFA  & \NL-\complete     & (Cor.~\ref{rponfasUnary})
                & \coNP-\complete   & (Cor.~\ref{thmMainRPONFASfixed})
                & \PSpace-\complete & (Thm.~\ref{thmMainRPONFAS})\\
        poNFA   & \NL-\complete     & (Thm.~\ref{ponfasUnary})
                & \PSpace-\complete & (Thm.~\ref{thmMainPONFAS})
                & \PSpace-\complete & \cite{AhoHU74} \\
        NFA     & \coNP-\complete   & \cite{StockmeyerM73}
                & \PSpace-\complete & \cite{AhoHU74} 
                & \PSpace-\complete & \cite{AhoHU74}
      \end{tabular}%
    \hfill\mbox{}%
    \caption{Complexity of deciding universality}\label{table_results}
  \end{table}
  However, this reduction of complexity is not preserved for unrestricted alphabets. We use a novel construction of rpoNFAs that characterize certain exponentially long words to show that universality is \PSpace-complete even for rpoNFAs if the alphabet may grow polynomially. Our complexity results are summarized in Table~\ref{table_results}.

  As a by-product, we show that rpoNFAs provide another characterization of $\R$-trivial languages introduced and studied by Brzozowski and Fich~\cite{BrzozowskiF80}, and we establish the complexity of detecting $\R$-triviality and $k$-$\R$-triviality for rpoNFAs.

  From the practical point of view, the problems of inclusion and equivalence of two languages, which are closely related to universality, are of interest, e.g., in optimization. Indeed, universality can be expressed either as the inclusion $\Sigma^* \subseteq L$ or as the equivalence $\Sigma^* = L$. Although equivalence can be seen as two inclusions, the complexity of inclusion does not play the role of a lower bound. For instance, for two deterministic context-free languages inclusion is undecidable~\cite{Friedman76}, whereas equivalence is decidable~\cite{Senizergues97}. However, the complexity of universality gives a lower bound on the complexity of both inclusion and equivalence, and we show that, for the partially ordered NFAs studied in this paper, the complexities of inclusion and equivalence coincide with the complexity of universality.
  
  This paper is a full version of the work~\cite{mfcs16:mktmmt} presented at the 41st International Symposium on Mathematical Foundations of Computer Science.

\section{Preliminaries and Definitions}\label{sec2}
  We assume that the reader is familiar with automata theory~\cite{AhoHU74}. The cardinality of a set $A$ is denoted by $|A|$ and the power set of $A$ by $2^A$. An {\em alphabet\/} $\Sigma$ is a finite nonempty set. A {\em word\/} over $\Sigma$ is any element of the free monoid $\Sigma^*$, the {\em empty word\/} is denoted by $\eps$. A {\em language\/} over $\Sigma$ is a subset of $\Sigma^*$. For a language $L$ over $\Sigma$, let $\overline{L}=\Sigma^*\setminus L$ denote its complement.
  
  A {\em subword\/} of $w$ is a word $u$ such that $w = w_1 u w_2$, for some words $w_1,w_2$; $u$ is a {\em prefix\/} of $w$ if $w_1=\eps$ and it is a {\em suffix\/} of $w$ if $w_2=\eps$.
  
  A {\em nondeterministic finite automaton\/} (NFA) is a quintuple $\A = (Q,\Sigma,\cdot,I,F)$, where $Q$ is a finite nonempty set of states, $\Sigma$ is an input alphabet, $I\subseteq Q$ is a set of initial states, $F\subseteq Q$ is a set of accepting states, and $\cdot \colon Q\times\Sigma \to 2^Q$ is the transition function that can be extended to the domain $2^Q \times \Sigma^*$ by induction. The language {\em accepted\/} by $\A$ is the set $L(\A) = \{w\in\Sigma^* \mid I \cdot w \cap F \neq \emptyset\}$. We often omit $\cdot$ and write simply $Iw$ instead of $I \cdot w$. The NFA $\A$ is {\em complete\/} if for every state $q$ and every letter $a$ in $\Sigma$, the set $q\cdot a$ is nonempty. It is {\em deterministic\/} (DFA) if $|I|=1$ and $|q\cdot a|=1$ for every state $q$ in $Q$ and every letter $a$ in $\Sigma$. 

  A {\em path\/} $\pi$ from a state $q_0$ to a state $q_n$ under a word $a_1a_2\cdots a_{n}$, for some $n\ge 0$, is a sequence of states and input symbols $q_0 a_1 q_1 a_2 \cdots q_{n-1} a_{n} q_n$ such that $q_{i+1} \in q_i\cdot a_{i+1}$, for $i=0,1,\ldots,n-1$. Path $\pi$ is {\em accepting\/} if $q_0\in I$ and $q_n\in F$. 
  A path is {\em simple\/} if all the states are pairwise distinct.

  A {\em deterministic Turing machine} (DTM) is a tuple $M = \tuple{Q,T,\tmInputAlphabet,\gamma,\blank,q_o,q_f}$, where $Q$ is the finite state set, $T$ is the tape alphabet, $I\subseteq T$ is the input alphabet, $\blank\in T \setminus I$ is the blank symbol, $q_o$ is the initial state, $q_f$ is the accepting state, and $\gamma$ is the transition function mapping $Q\times T$ to $Q\times T \times \{L,R,S\}$, see Aho et al.~\cite{AhoHU74} for details.
  
  The {\em universality problem\/} asks, given an automaton $\A$ over $\Sigma$, whether $L(\A)=\Sigma^*$. The {\em inclusion problem\/} asks, given two automata $\A$ and $\B$ over a common alphabet, whether $L(\A) \subseteq L(\B)$, and the {\em equivalence problem\/} asks whether $L(\A) = L(\B)$.

\section{Partially Ordered NFAs}\label{sec:ponfas}

  In this section, we introduce poNFAs, recall their characterization in terms of the Straubing-Th\'erien hierarchy, and show that universality remains \PSpace-complete even when restricting to binary alphabets. Merely the case of unary alphabets turns out to be simpler.

  \begin{definition}\label{poNFA} 
    Let $\A$ be an NFA. A state $q$ is \emph{reachable} from a state $p$, written $p\leq q$, if there exists a word $w\in\Sigma^*$ such that $q\in p\cdot w$. We write $p< q$ if $p\leq q$ and $p\neq q$. $\A$ is a \emph{partially ordered NFA} (\emph{poNFA}) if $\le$ is a partial order.
  \end{definition}
  
  The expressive power of poNFAs can be characterized by the \emph{Straubing-Th\'erien (ST) hierarchy} \cite{Straubing81,Therien81}. For an alphabet $\Sigma$, level 0 of this hierarchy is defined as $\mathscr{L}(0)=\{\emptyset, \Sigma^*\}$. For integers $n\geq 0$, the levels $\mathscr{L}(n+\frac{1}{2})$ and $\mathscr{L}(n+1)$ are as follows:
  
  \begin{itemize}
    \item $\mathscr{L}(n+\frac{1}{2})$ consists of all finite unions of languages $L_0 a_1 L_1 a_2 \cdots a_k L_k$, with $k\geq 0$, $L_0,\ldots, L_k\in\mathscr{L}(n)$, and $a_1,\ldots,a_k\in\Sigma$;
    \item $\mathscr{L}(n+1)$ consists of all finite Boolean combinations of languages from level $\mathscr{L}(n+\frac{1}{2})$.
  \end{itemize}
  
  Note that the levels of the hierarchy contain only \emph{star-free} languages by definition. It is known that the hierarchy does not collapse on any level \cite{BrzozowskiK78}, but the problem of deciding if a language belongs to some level $k$ is largely open for $k>\frac{7}{2}$ \cite{AlmeidaBKK15,Place15,PlaceZ15}. The ST hierarchy further has close relations to the \emph{dot-depth hierarchy}~\cite{BrzozowskiK78,CohenB71,Straubing85} and to complexity theory~\cite{Wagner04}.

  Interestingly, the languages recognized by poNFAs are exactly the languages on level~$\frac{3}{2}$ of the Straubing-Th\'erien hierarchy \cite{SchwentickTV01}. Since the hierarchy is proper, this means that poNFAs can only recognize a strict subset of star-free regular languages.  In spite of this rather low expressive power, the universality problem of poNFAs has the same worst-case complexity as for general NFAs, even when restricting to a fixed alphabet with only a few letters. 
  
  \begin{theorem}\label{thmMainPONFAS}
    For every alphabet $\Sigma$ with $|\Sigma|\geq 2$, the universality problem for poNFAs over $\Sigma$ is \PSpace-complete.
  \end{theorem}
  \begin{pf}
    Membership follows from the fact that universality is in \PSpace for NFAs~\cite{GareyJ79}.
    
    To show hardness, we modify the construction of Aho et al.~\cite[Section~10.6]{AhoHU74} to work on a two-letter alphabet. Consider a polynomial $p$ and a $p$-space-bounded DTM $M = \tuple{Q,T,\tmInputAlphabet,\gamma,\blank,q_o,q_f}$. Without loss of generality, we assume that $q_o\neq q_f$. We define an encoding of runs of $M$ as a word over a given alphabet. For any input $x\in\tmInputAlphabet^*$, we construct, in polynomial time, a regular expression $R_x$ that represents all words that do \emph{not} encode an accepting run of $M$ on $x$. Therefore, $R_x$ matches all words if and only if $M$ does not accept $x$. The claim then follows by showing that $R_x$ can be encoded by a poNFA.

    A configuration of $M$ on an input $x$ consists of a current state $q\in Q$, the position $1\leq \ell\leq p(|x|)$ of the read/write head, and the current tape contents $\theta_1,\ldots,\theta_{p(|x|)}$ with $\theta_i\in T$. We represent it by a sequence 
    \[
      \tuple{\theta_1,\eps}\cdots\tuple{\theta_{\ell-1},\eps}\tuple{\theta_{\ell},q}\tuple{\theta_{\ell+1},\eps}\cdots\tuple{\theta_{p(|x|)},\eps}
    \]
    of symbols from $T\times(Q\cup\{\eps\})$. We denote $T\times(Q\cup\{\eps\})$ by $\Delta$. A potential run of $M$ on $x$ is represented by word $\# w_1 \# w_2 \# \cdots \# w_m \#$, where $w_i\in\Delta^{p(|x|)}$ and $\#\notin\Delta$ is a fresh separator symbol. One can construct a regular expression recognizing all words over $\Delta\cup\{\#\}$ that do not correctly encode a run of $M$ at all, or that encode a run that is not accepting \cite{AhoHU74}.
        
    We encode symbols of $\Delta\cup\{\#\}$ using the fixed alphabet $\Sigma = \{0,1\}$. For each $\delta\in\Delta\cup\{\#\}$, let $\hat{\delta}_1\cdots\hat{\delta}_K\in \{0,1\}^K$ be the unique binary encoding of length $K =\left\lceil\log_2\left(|\Delta\cup\{\#\}|\right)\right\rceil$. We define $\enc(\delta)$ to be the binary sequence
    \[
      \enc(\delta) = 001\hat{\delta}_11\hat{\delta}_21\cdots\hat{\delta}_K1
    \]
    of length $L=2K+3$. We extend $\enc$ to words and sets of symbols as usual:    $\enc(\delta_1\cdots\delta_m)=\enc(\delta_1)\cdots\enc(\delta_m)$ and $\enc(\Delta')=\{\enc(\delta)\mid\delta\in\Delta'\}$. Importantly, any word of the form $\enc(\delta_1\cdots\delta_m)$ contains $00$ only at positions that are multiples of $L$, marking the start of one encoded symbol.
    
    We now construct the regular expression $R_x$ that matches all words of $\Sigma^*$ that do not represent an accepting computation of $M$ on $x$. We proceed in four steps: 
    \begin{description}
      \item[(A)] We detect all words that contain words from $\Sigma^*$ that are not of the form $\enc(\delta)$; 
      
      \item[(B)] We detect all words that do not start with the initial configuration; 
      
      \item[(C)] We detect all words that do not encode a valid run since they violate a transition rule; and 
      
      \item[(D)] We detect all words that encode non-accepting runs, or runs that end prematurely.
    \end{description}
    
    For (A), note that a word $w\in\Sigma^*$ that is not of the form $\enc(v)$ for any word $v\in(\Delta\cup\{\#\})^*$ must either
    (A.1) start with $1$ or $01$;
    (A.2) end with $0$;
    (A.3) contain a word $00\Sigma^{L-2}$ that is not in $\enc(\Delta\cup\{\#\})$;
    (A.4) contain a word from $\enc(\Delta\cup\{\#\})\{1,01\}$; or
    (A.5) end in a word $00\Sigma^{M}$ with $M< L-2$.
    Using $E$ to abbreviate $\enc(\Delta\cup\{\#\})$ and $\bar{E}$ to abbreviate $00\Sigma^{L-2}\setminus E$ (both sets of polynomially many binary sequences), we can express (A.1)--(A.5) in the regular expression
    \begin{equation}
      (1\Sigma^* + 01\Sigma^*) +
      (\Sigma^* 0) +
      \left(\Sigma^*\bar{E}\Sigma^*\right) +
      \left(\Sigma^*E(1+01)\Sigma^*\right) +
      \left(\Sigma^* 00 (\Sigma + \Sigma^2 + \cdots + \Sigma^{L-3})\right)
      \label{eq_re_wrong_enc}
    \end{equation}
    where we use finite sets $\{e_1,\ldots,e_m\}$ to denote regular expressions $(e_1+\cdots +e_m)$, as usual. All sets in \eqref{eq_re_wrong_enc} are polynomial in size, so that the overall expression is polynomial. The expression \eqref{eq_re_wrong_enc} can be captured by a poNFA since the only cycles required arise when translating $\Sigma^*$; they can be expressed as self-loops. All other repetitions of the form $\Sigma^i$ in \eqref{eq_re_wrong_enc} can be expanded to polynomial-length sequences without cycles.
    
    For (B), we want to detect all words that do not start with the word
    \[
      w = \enc(\#\tuple{x_1,q_0}\allowbreak\tuple{x_2,\eps}\cdots\allowbreak\tuple{x_{|x|},\eps}\tuple{\blank,\eps}\cdots\tuple{\blank,\eps}\#) = \enc(v_0 v_1 \cdots v_{p(x)+1})
    \]
    of length $(p(|x|)+2)L$. This happens if (B.1) the word is shorter than $(p(|x|)+2)L$, or (B.2), starting at position $jL$ for $0\leq j\leq p(|x|)+1$, there is a word from the polynomial set $\Sigma^L\setminus\{\enc(v_j)\}$, which we abbreviate by $\bar{E}_j$. We can capture (B.1) and (B.2) in the regular expression
    \begin{equation}
      \left(\varepsilon + \Sigma + \Sigma^2 +\cdots+ \Sigma^{L(p(|x|)+2)-1}\right) 
      + \sum_{0\leq j\leq p(|x|)+1} (\Sigma^{jL} \cdot \bar{E}_j\cdot\Sigma^*)\label{eq_re_wrong_start}
    \end{equation}
    The empty expression $\varepsilon$ is used for readability; it can easily be expressed in the NFA encoding. As before, it is easy to see that this expression is polynomial and does not require any nontrivial cycles when encoded in an NFA. Note that we ensure that the surrounding $\#$ in the initial configuration are present.
    
    For (C), we need to check for incorrect transitions. Consider again the encoding $\#w_1\#\cdots\#w_m\#$ of a sequence of configurations with a word over $\Delta\cup\{\#\}$, where we can assume that $w_1$ encodes the initial configuration according to (A) and (B). In an encoding of a valid run, the symbol at any position $j\geq p(|x|)+2$ is uniquely determined by the symbols	at positions $j-p(|x|)-2$, $j-p(|x|)-1$, and $j-p(|x|)$, corresponding to the cell and its left and right neighbor in the previous configuration. Given symbols $\delta_\ell,\delta,\delta_r\in\Delta\cup\{\#\}$, we can therefore define $f(\delta_\ell,\delta,\delta_r)\in\Delta\cup\{\#\}$ to be the symbol required in the next configuration. The case where $\delta_\ell=\#$ or $\delta_r=\#$ corresponds to transitions applied at the left and right edge of the tape, respectively; for the case that $\delta=\#$, we define $f(\delta_\ell,\delta,\delta_r)=\#$, ensuring that the separator $\#$ is always present in successor configurations as well. We can then check for invalid transitions using the regular expression
    \begin{equation}
      \sum_{\delta_\ell,\delta,\delta_r \in \Delta\cup\{\#\}} \Sigma^*\cdot \enc(\delta_\ell\delta\delta_r)\cdot  \Sigma^{L(p(|x|)-1)} \cdot \enc(\overline{f}(\delta_\ell,\delta,\delta_r))\cdot \Sigma^*
      \label{eq_re_wrong_trans}
    \end{equation}
    where $\overline{f}(\delta_\ell,\delta,\delta_r) = \Delta\cup\{\#\}\setminus\{f(\delta_\ell,\delta,\delta_r)\}$. Polynomiality and poNFA-expressibility are again immediate. Note that expression \eqref{eq_re_wrong_trans} only detects wrong transitions if a (long enough) next configuration exists. The case that the run stops prematurely is covered next.
    
    Finally, for (D) we detect all words that either (D.1) end in a configuration that is incomplete (too short) or (D.2) end in a configuration that is not in the final state $q_f$. Abbreviating $\enc(T\times(Q\setminus\{q_f\}))$ as $\bar{E}_f$, and using similar ideas as above, we obtain
    \begin{align}
      \left( \Sigma^* \enc(\#) (\Sigma^L + \cdots + \Sigma^{p(|x|)L}) \right) +
      \left( \Sigma^* \bar{E}_f (\varepsilon + \Sigma^L + \cdots + \Sigma^{(p(|x|)-1)L}) \enc(\#)  \right)
      \label{eq_re_wrong_final}
    \end{align}
    and this can again be expressed as a polynomial poNFA.
    
    The expressions \eqref{eq_re_wrong_enc}--\eqref{eq_re_wrong_final} together then detect all non-accepting or wrongly encoded runs of $M$. In particular, if we start from the correct initial configuration (\eqref{eq_re_wrong_start} does not match), then for \eqref{eq_re_wrong_trans} not to match, all complete future configurations must have exactly one state and be delimited by encodings of $\#$. Expressing the regular expressions as a single poNFA of polynomial size, we have thus reduced the word problem of polynomially space-bounded Turing machines to the universality problem of poNFAs.
  \qed\end{pf}

  Ellul et al.~\cite[Section~5]{EllulKSW05} give an example of a regular expression over a 5-letter alphabet such that the shortest non-accepted word is of exponential length, and which can also be encoded as a poNFA. Our previous proof shows such an example for an alphabet of two letters, if we use a Turing machine that runs for exponentially many steps before accepting. Note, however, that this property alone would not imply Theorem~\ref{thmMainPONFAS}.

\subsection*{Unary Alphabet}
  Reducing the size of the alphabet to one leads to a reduction in complexity. This is expected, since the universality problem for NFAs over a unary alphabet is merely \coNP-complete \cite{StockmeyerM73}. For poNFAs, the situation is even simpler:
  
  \begin{theorem}\label{ponfasUnary}
    The universality problem for poNFAs over a unary alphabet is \NL-complete. It can be checked in linear time.
  \end{theorem}
  \begin{pf}
    Let $\A$ be a poNFA over the alphabet $\{a\}$, and let $n$ be the number of states in $\A$. Language $L(\A)$ is infinite if and only if a word of length $n$ is accepted by $\A$. If $a^{n}$ is accepted, then there must be a simple path from an initial state to an accepting state via a state with a self-loop. Therefore, all words of length $n$ or more are accepted. It remains to check that $\eps,a,\ldots,a^n$ are accepted, which amounts to $n$ acceptance checks that can be realized in nondeterministic logarithmic space. Notice that, using linear space, these checks altogether can be done in linear time.
    Hardness can be shown by reducing the \NL-complete DAG-reachability problem~\cite{Jones75}. Let $G$ be a directed acyclic graph, and let $s$ and $t$ be two nodes of $G$. We define a poNFA $\A$ as follows. With each node of $G$, we associate a state in $\A$. Whenever there is an edge from $i$ to $j$ in $G$, we add an $a$-transition from $i$ to $j$ in $\A$. We add a self-loop labeled by $a$ to $t$. The initial state of $\A$ is state $s$, all states are final. Then $\A$ is universal if and only if there is a path from $s$ to $t$ in $G$.
  \qed\end{pf}

\section{Restricted Partially Ordered NFAs}\label{sec:rponfas} 
  We now introduce restricted poNFAs, which are distinguished by deterministic self-loops.
  We relate them to the known class of $\R$-trivial languages, and we establish complexity results for
  deciding if a language falls into this class.
  
  \begin{definition}\label{rpoNFA}
    A \emph{restricted partially ordered NFA (rpoNFA)} is a poNFA such that, for every state $q$ and
    symbol $a$, if $q\in q\cdot a$ then $q\cdot a = \{q\}$.
  \end{definition}

  We will show below that rpoNFAs characterize $\R$-trivial languages \cite{BrzozowskiF80}.
  To introduce this class of languages, we first require some auxiliary definitions.
  A word $v=a_1 a_2 \cdots a_n$ is a \emph{subsequence} of a word $w$,
  denoted by $v \preccurlyeq w$, if $w\in \Sigma^* a_1 \Sigma^* a_2 \Sigma^* \cdots \Sigma^* a_n \Sigma^*$.
  For $k\ge 0$, we write $\sub_k(v) =\{u\in\Sigma^* \mid u\preccurlyeq v,\, |u|\le k\}$ for the set of
  all subsequences of $v$ of length up to $k$.
  Two words $w_1, w_2$ are \emph{$\sim_k$-equivalent}, written $w_1 \sim_k w_2$,
  if $\sub_k(w_1)=\sub_k(w_2)$. Then $\sim_k$ is a congruence (for concatenation) of
  finite index (i.e., with finitely many equivalence classes)~\cite{Simon1972}.
  $\R$-trivial languages are defined by defining a related congruence $\sim^{\R}_{k}$
  that considers subsequences of prefixes:
  
  \begin{definition}
    Let $x,y\in\Sigma^*$ and $k\ge 0$. Then $x \sim^{\R}_{k} y$ if and only if
    
    \begin{itemize}
      \item for each prefix $u$ of $x$, there exists a prefix $v$ of $y$ such that $u \sim_k v$, and
      \item for each prefix $v$ of $y$, there exists a prefix $u$ of $x$ such that $u \sim_k v$.
    \end{itemize}
    
    A regular language is \emph{$k$-$\R$-trivial} if it is a union of $\sim^{\R}_{k}$ classes,
    and it is \emph{$\R$-trivial} if it is $k$-$\R$-trivial for some $k\ge 0$.
  \end{definition}
  
  It is known that $x \sim^{\R}_{k} y$ implies $x \sim_k y$ and (if $k\geq 1$)
  $x \sim^{\R}_{k-1} y$ \cite{BrzozowskiF80}.
  Therefore, every $k$-$\R$-trivial language is also $(k+1)$-$\R$-trivial.
  Moreover, it has been shown that a language $L$ is $\R$-trivial if and only if
  the minimal DFA recognizing $L$ is partially ordered~\cite{BrzozowskiF80}. We can lift this result to
  characterize the expressive power of rpoNFAs.
  
  \begin{theorem}\label{rpoNFAsRlangs}
      A regular language is $\R$-trivial if and only if it is accepted by an rpoNFA.
  \end{theorem}
    \begin{pf}
    Brzozowski and Fich~\cite{BrzozowskiF80} have shown that every $\R$-trivial language is accepted by a partially ordered DFA. As a partially ordered DFA is an rpoNFA, this concludes this direction.
    
    To prove the other direction, notice that every rpoNFA can be decomposed into a finite number of DFAs. More specifically, let $\A$ over $\Sigma$ be an rpoNFA. For a state $q$, let $\Sigma_q=\{ a \in \Sigma \mid q \in q \cdot a\}$ be the set of all symbols that appear in self-loops in state $q$. Let $q_1 a_1 q_2 a_2 \cdots q_n a_n q_{n+1}$ be a simple accepting path in $\A$. Then it defines an expression $\Sigma_{q_1}^* a_1 \Sigma_{q_2}^* a_2 \cdots \Sigma_{q_n}^* a_n \Sigma_{q_{n+1}}^*$ with the property $a_i \notin \Sigma_{q_i}$ for $1\le i \le n$. Since every NFA has only finitely many simple paths, the proof now follows from the results of Brzozowski and Fich~\cite{BrzozowskiF80}, who have shown that a language is $\R$-trivial if and only if it is a finite union of $\R$-expressions, i.e., expressions of the form $\Sigma_1^* a_1 \Sigma_2^* a_2 \cdots \Sigma_m^* a_m \Sigma_{m+1}^*$, for some $m\ge 0$, where $a_i\notin \Sigma_i$ for $1\le i \le m$.
  \qed\end{pf}
  
  This characterization in terms of automata with forbidden patterns can be compared to
  results of Gla\ss{}er and Schmitz, who use DFAs with a forbidden pattern to obtain
  a characterization of level~$\frac{3}{2}$ of the dot-depth hierarchy~\cite{GlasserS08,Schmitz}.
  
  We can further relate the \emph{depth} of rpoNFAs to $k$-$\R$-trivial languages.
  Recall that the depth of an rpoNFA $\A$, denoted by $\depth(\A)$, is the
  number of input symbols on a longest simple path of $\A$ that starts in an initial state.
 
  \begin{theorem}\label{thmMain}
    The language recognized by a complete rpoNFA $\A$ is $\depth(\A)$-$\R$-trivial.
  \end{theorem}
  
  The proof of Theorem~\ref{thmMain} follows from Lemmas~\ref{lemmaAi} and~\ref{lemmaMain} proved below. 
  
  Let $p$ be a state of an NFA $\A=(Q,\Sigma,\cdot,I,F)$. The {\em sub-automaton\/} of $\A$ induced by state $p$ is the automaton $\A_p = (\reach(p),\Sigma,\cdot_p,\{p\},F\cap \reach(p))$ with state $p$ being the sole initial state and with only those states of $\A$ that are reachable from $p$; formally, $\reach(p)$ denotes the set of all states reachable from state $p$ in $\A$ and $\cdot_p$ is the restriction of $\cdot$ to $\reach(p)\times\Sigma$.

  The following lemma is clear.
  
  \begin{lemma}\label{lemmaAi}
    Let $\A$ be an rpoNFA with $I$ denoting the set of initial states. Then the language $L(\A) = \bigcup_{i\in I} L(\A_i)$, where every sub-automaton $\A_i$ is an rpoNFA.
  \end{lemma}
  
  Thus, it is sufficient to prove the theorem for rpoNFAs with a single initial state. Indeed, if $\A_i$ is of depth $k_i$, then its language is $k_i$-$\R$-trivial by Lemma~\ref{lemmaMain}. Since every $k$-$\R$-trivial language is also $(k+1)$-$\R$-trivial, the union of $L(\A_i)$ is $\max\{k_i \mid i \in I\}$-$\R$-trivial.

  We need the following two lemmas first. For a word $w$, we denote by $\alp(w)$ the set of all letters occurring in $w$.
  
  \begin{lemma}[\cite{KlimaP13}]\label{lemma1}
    Let $\ell \ge 1$, and let $x, y \in \Sigma^*$ be such that $x \sim_{\ell} y$. Let $x = x' a x''$ and $y = y' a y''$ such that $a\notin \alp(x' y')$. Then $x'' \sim_{\ell-1} y''$.
  \end{lemma}
  
  \begin{lemma}\label{lemma2}
    Let $\ell \ge 1$, and let $x, y \in \Sigma^*$ be such that $x \sim^{\R}_{\ell} y$. Let $x = x' a x''$ and $y = y' a y''$ such that $a\notin \alp(x' y')$. Then $x'' \sim^{\R}_{\ell-1} y''$.
  \end{lemma}
  \begin{pf}
    Let $u''$ be a prefix of $x''$. Consider the prefix $ u = x'au''$ of $x$. Since $x\sim^{\R}_{\ell} y$, there exists a prefix $v$ of $y$ such that $u \sim_\ell v$. Then $\ell \ge 1$ implies that letter $a$ appears in $v$. Thus, we can write $v=y'av''$. By Lemma~\ref{lemma1}, $u'' \sim_{\ell-1} v''$. Thus, for any prefix $u''$ of $x''$, there exists a prefix $v''$ of $y''$ such that $u'' \sim_{\ell-1} v''$. Similarly the other way round, and therefore $x'' \sim^{\R}_{\ell-1} y''$.
  \qed\end{pf}

  \begin{lemma}\label{lemmaMain}
    Let $\A$ be a complete rpoNFA with a single initial state and depth $k$. Then the language $L(\A)$ is $k$-$\R$-trivial.
  \end{lemma}
  \begin{pf}
    Let $\A=(Q,\Sigma,\cdot,i,F)$. If the depth of $\A$ is 0, then $L(\A)$ is either $\emptyset$ or $\Sigma^*$, which are both $0$-$\R$-trivial by definition. Thus, assume that the depth of $\A$ is $\ell \ge 1$ and that the claim holds for rpoNFAs of depth less than $\ell$. Let $u, v\in \Sigma^*$ be such that $u \sim^{\R}_{\ell} v$. We prove that $u$ is accepted by $\A$ if and only if $v$ is accepted by $\A$. 
    
    Assume that the word $u$ is accepted by $\A$ and fix an accepting path of $u$ in $\A$. Let $\Sigma_i = \{ a\in \Sigma \mid i \in i \cdot a \}$ denote the set of all letters under which there is a self-loop in state $i$. If $\alp(u)\subseteq \Sigma_i$, then the definition of rpoNFA $\A$ implies that $i\in F$. Since $\ell \ge 1$ implies that $\alp(u) = \alp(v)$, we have that $v$ is also accepted in state $i$.
    
    If $\alp(u)\not\subseteq\Sigma_i$, then 
    \[
      u = u'au'' \quad\text{ and }\quad v = v'bv''
    \]
    where $u',v'\in\Sigma_i^*$, $a,b\in \Sigma\setminus\Sigma_i$, and $u'',v''\in \Sigma^*$. Let $p\in i\cdot a$ be a state on the fixed accepting path of $u$, and let $\A_p$ be the sub-automaton of $\A$ induced by state $p$. Notice that $\A_p$ is a complete rpoNFA of depth at most $\ell -1 $, and that $\A_p$ accepts $u''$.

    If $a\neq b$, then $u = u' a u_0 b u_1$ and $v = v' b v_0 a v_1$, where the depicted $a$ and $b$ are the first occurrences of those letters from the left, that is, $b \notin \alp(u' a u_0)\cup\alp(v')$ and $a \notin \alp(u')\cup\alp(v' b v_0)$. 
    If $\ell = 1$, let $z = u'a$ be a prefix of $u$. Since $u \sim^{\R}_{1} v$, there exists a prefix $t$ of $v$ such that $z \sim_1 t$. Because $a \in \alp(z)$, we also have that $a \in \alp(t)$, which implies that $t = v' b v_0 a t'$, for some $t'$ being a prefix of $v_1$. But then $b \in \alp(t) \setminus \alp(z)$, which is a contradiction with $z \sim_1 t$. If $\ell \ge 2$, let $z = u' a u_0 b$ be a prefix of $u$. Since $u \sim^{\R}_{\ell} v$, there exists a prefix $t$ of $v$ such that $z \sim_\ell t$. Because $a,b \in \alp(z)$, we also have that $a,b \in \alp(t)$, which implies that $t = v' b v_0 a t'$, for some $t'$ being a prefix of $v_1$. But then $ba \in \sub_\ell(t) \setminus \sub_\ell(z)$, which is a contradiction with $z \sim_\ell t$. Thus, $u \sim^{\R}_{\ell} v$ implies that $a=b$. 
    
    If $a = b$, Lemma~\ref{lemma2} implies that $u'' \sim^{\R}_{\ell-1} v''$. By the induction hypothesis, $u''$ is accepted by $\A_p$ if and only if $v''$ is accepted by $\A_p$. Hence, $v=v'av''$ is accepted by $\A$, which was to be shown.
  \qed\end{pf}
  
\begin{pf}[of Theorem \ref{thmMain}]
  By Lemma \ref{lemmaAi} and the definition of $k$-$\mathcal{R}$-triviality, the language recognized by the rpoNFA $\mathcal{A}$ is $\depth(\A)$-$\R$-trivial if the language recognized by each $\mathcal{A}_i$ is $\depth(\A)$-$\R$-trivial. Since the depth of every $\mathcal{A}_i$ is at most the depth of $\mathcal{A}$, Lemma \ref{lemmaMain} concludes the proof.
\qed\end{pf}

	Similar relationships have been studied for $\J$-trivial languages~\cite{KlimaP13,ptnfas}, but we are
	not aware of any such investigation for $\R$-trivial languages.

	Finally, we may ask how difficult it is to decide whether a given NFA $\A$
	accepts a language that is $\R$-trivial or $k$-$\R$-trivial for a specific $k\geq 0$.
	For most levels of the ST hierarchy, it is not even known if this problem is decidable,
	and when it is, exact complexity bounds are often missing~\cite{PlaceZ15}. 
	The main exception are $\J$-trivial languages -- level 1 of the hierarchy -- which have recently
	attracted some attention, motivated by applications in algebra and XML databases~\cite{HofmanM15,KlimaP13,dlt15}.
		
	To the best of our knowledge, the following complexity results for recognizing ($k$-)$\R$-trivial languages had not been obtained previously.

  \begin{theorem}\label{RtrivNFA}
	Given an NFA $\A$, it is \PSpace-complete to decide if the language accepted by $\A$
	is $\R$-trivial.
  \end{theorem}
  \begin{pf}
    The hardness follows from Theorem~3.1 in Hunt~III and Rosenkrantz~\cite{HuntR78}.
    To decide whether the language $L(\A)$ is $\R$-trivial means to check whether its equivalent (minimal) DFA is partially ordered. The non-partial-order of the DFA can be checked in \PSpace by nondeterministically guessing two reachable subsets of states and verifying that they are inequivalent and reachable from each other.
    This shows that $\R$-triviality is \PSpace-complete.
  \qed\end{pf}

  To prove a similar claim for $k$-$\R$-triviality, we use some results from the literature.

  \begin{lemma}[\cite{BrzozowskiF80}]\label{lemmaMinWord}
    Every congruence class of $\sim^{\R}_{k}$ contains a unique element of minimal length. If $a_1,a_2, \ldots, a_n \in \Sigma$, then $a_1a_2\cdots a_n$ is minimal if and only if $\sub_k(\eps) \subsetneq \sub_k(a_1) \subsetneq \sub_k(a_1a_2) \subsetneq \cdots \subsetneq \sub_k(a_1a_2\cdots a_n)$.
  \end{lemma}

  The maximal length of such a word has also been studied~\cite{dlt15}.
  \begin{lemma}[\cite{dlt15}]\label{lemmaDLT}
    Let $\Sigma$ be an alphabet of cardinality $|\Sigma|\ge 1$, and let $k\ge 1$. The length of a longest word $w$ such that $\sub_k(w) = \{ v \in\Sigma^* \mid |v| \le k\}$, and, for any two distinct prefixes $w_1$ and $w_2$ of $w$, $\sub_k(w_1)\neq \sub_k(w_2)$, is exactly $\binom{k+|\Sigma|}{k} - 1$.
  \end{lemma}

  Lemmas~\ref{lemmaMinWord} and \ref{lemmaDLT} provide the main ingredients for
  showing membership in \PSpace.
  
  \begin{theorem}\label{kRtrivNFA}
    Given an NFA $\A$ and $k\ge 0$, it is \PSpace-complete to decide if the language accepted by $\A$ is $k$-$\R$-trivial.
  \end{theorem}
  \begin{pf}
    Again, the hardness follows from Theorem~3.1 in Hunt~III and Rosenkrantz~\cite{HuntR78}.

    To prove the membership, let $\A$ be an NFA over an $n$-letter alphabet $\Sigma$. By definition, every $k$-$\R$-trivial language is a finite union of $\sim^{\R}_{k}$-classes. By Lemmas~\ref{lemmaMinWord} and~\ref{lemmaDLT}, every class has a unique shortest representative of length at most $\binom{k+n}{k}-1$. Since $k$ is a constant, this number is polynomial, $O(n^k)$. If $L(\A)$ is not $k$-$\R$-trivial, then there exists a class $C_w = w/_{\sim^{\R}_{k}}$, where $w$ is the unique shortest representative, such that $C_w \cap L(\A) \neq \emptyset$ and $C_w \cap \overline{L(\A)} \neq \emptyset$. The nondeterministic algorithm can guess $w$ and build the minimal DFA accepting the class $C_w$ as described below. Having this, the intersections can be checked in \PSpace. (The non-emptiness of the intersection with a complemented NFA can be verified, for instance, by the on-the-fly determinization of the NFA and reverting the status of the reached state, or by building and alternating finite automaton and checking non-emptiness~\cite{HolzerK11}).
    
    We construct the minimal incomplete DFA $\D_w$ recognizing only the word $w$. It consists of $|w|+1$ states labeled by prefixes of $w$ so that the initial state is labeled with $[\eps]$ and the only accepting state is labeled with $[w]$. The transitions are defined so that if $w=uau'$, then $[u] \cdot a = [ua]$. Now, for every prefix $v$ of $w$ and every letter $b$ such that $\sub_k(v) = \sub_k(vb)$, we add the self-loop $[v]\cdot b = [v]$ to $\D_w$. Notice that for $w=uau'$, $\sub_k(u) \neq \sub_k(ua)$ by the properties of the unique shortest representative, c.f.~Lemma~\ref{lemmaMinWord}, and therefore the construction produces a DFA. We make it complete by adding a sink state, if needed. Denote the obtained DFA by $\D$. We claim that $L(\D) = C_w$.

    \begin{claim}\label{claim1}
      $L(\D) \subseteq C_w$.
    \end{claim}
    \begin{pf}
      Let $w' \in L(\D)$. We show that $w' \sim^{\R}_{k} w$. To do this, let $w = a_1 a_2 \cdots a_n$. Then, by the structure of $\D$, $w' = u_0 a_1 u_1 a_2 u_2 \cdots u_{n-1} a_n u_n$, for some words $u_i$ that are read in self-loops of states $[a_1a_2\cdots a_i]$, for $0\le i \le n$. 
      
      By definition of $\sim^{\R}_{k}$, we need to show that for each prefix $u$ of $w'$, there exists a prefix $v$ of $w$ such that $u \sim_k v$, and that for each prefix $v$ of $w$, there exists a prefix $u$ of $w'$ such that $u \sim_k v$. We prove by induction on $i$, $0\le i\le n$, that $u_0 a_1 u_1 a_2 u_2 \cdots a_i u_i' \sim_k a_1a_2\cdots a_i$, where $u_i'$ is any prefix of $u_i$. 
      
      For $i=0$, we show that $\sub_k(u_0') = \sub_k(\eps)$ for any prefix $u_0'$ of $u_0$. Since $[\eps] \cdot u_0' = [\eps]$ in $\D$, we have that $\sub_k(u_0') = \sub_k(\eps)$. Indeed, if $u_0' = b_1 b_2 \cdots b_m$, then, by the construction of $\D$, $\eps \sim_k b_j$, for $1\le j \le m$. Since $\sim_k$ is a congruence, $\eps \sim_k b_1b_2\cdots b_m = u_0'$.

      Assume that it holds for $i-1$ and consider the prefixes $u_0 a_1 u_1 \cdots u_{i-1} a_i u_i'$ and $a_1\cdots a_{i-1} a_i$, where $u_i'$ is a prefix of $u_i$. By the induction hypothesis, $u_0 a_1 u_1 \cdots u_{i-1} \sim_k a_1\cdots a_{i-1}$, and by the congruence property of $\sim_k$, we obtain that $u_0 a_1 u_1 \cdots u_{i-1} a_i \sim_k a_1\cdots a_{i-1} a_i$. Let $u = u_0 a_1 u_1 \cdots u_{i-1} a_i$, $v=a_1\cdots a_{i-1} a_i$, and $u_i' = c_1 c_2 \cdots c_s$. By the construction of $\D$, the state $[v]$ has self-loops under all letters $c_j$, which means that $v \sim_k v c_j$, for $1 \le j \le s$. It implies that $v \sim_k v u_i'$, because $v c_{j+1} \cdots c_s \sim_k v c_j c_{j+1} \cdots c_s$ using $v \sim_k v c_j$ and the property that $\sim_k$ is a congruence. Since $u \sim_k v$ implies that $v u_i' \sim_k u u_i'$, we have that $v\sim_k v u_i' \sim_k u u_i'$, which was to be shown.
    \end{pf}
    
    \begin{claim}\label{claim2}
      $C_w \subseteq L(\D)$.
    \end{claim}
    \begin{pf}
      Let $w' \in \Sigma^*$ be such that $w' \sim^{\R}_{k} w$. We show that $w'$ is accepted by $\D$. For the sake of contradiction, assume that $w'$ does not belong to $L(\D)$. Let $w_1'$ denote the longest prefix of $w'$ that can be read by $\D$, that is, $w_1' = u_0 a_1 u_1 \cdots a_i u_i$, where $u_i$'s correspond to words read in self-loops and $a_i$ to letters of $w$. Let $w_1 = a_1 a_2\cdots a_i$ denote the corresponding prefix of $w$. Then $w = w_1 w_2$ and $w_1'w_2$ is accepted by $\D$. By Claim~\ref{claim1}, $w \sim^{\R}_{k} w_1' w_2$; namely, $w_1' = u_0 a_1 u_1 \cdots a_i u_i \sim_k a_1a_2\cdots a_i = w_1$. Since $w'$ is not accepted by $\D$, there is a letter $b$ such that $w_1' b$ is a prefix of $w'$ and it leads $\D$ to the sink state. Thus, $\sub_k(w_1) = \sub_k(w_1') \subsetneq \sub_k(w_1' b)$ by the construction of $\D$. Moreover, $w' \sim^{\R}_{k} w$ implies that there is a prefix $v$ of $w$ such that $w_1' b \sim_k v$. Since $\sub_k(w_1) \subsetneq \sub_k(w_1' b)$, there must be a letter $a$ such that $v = w_1 a y$, for some word $y$. Notice that $a\neq b$. Since $w$ is the unique minimal representative, $\sub_k(w_1) \subsetneq \sub_k(w_1 a)$, and hence there exists $x$ such that $x \in \sub_k(w_1 a)$ and $x \notin \sub_k(w_1) = \sub_k(w_1')$. Since $a\neq b$, $x \notin \sub_k(w_1' b)$, which is a contradiction with $w_1' b \sim_k v$.
    \end{pf}

    This completes the proof of Theorem~\ref{kRtrivNFA}.
  \qed\end{pf}

  In both previous theorems, hardness is shown by reduction from the universality problem for NFAs~\cite{AhoHU74,MeyerS72}. Hence it holds even for binary alphabets.
  For a unary alphabet, we can obtain the following result.
  
  \begin{theorem}\label{RtrivUnary}
    Given an NFA $\A$ over a unary alphabet, the problems of deciding if the language accepted by $\A$ is $\R$-trivial, or $k$-$\R$-trivial for a given $k\geq 0$, are both \coNP-complete.
  \end{theorem}
  \begin{pf}
    To show that $\R$-triviality for NFAs over a unary alphabet $\{a\}$ is in \coNP, we show that non-$\R$-triviality is in NP. It requires to check that the corresponding DFA is not partially ordered, which is if and only if there are $0\le \ell_1 < \ell_2 < \ell_3\le 2^n$, where $n$ is the number of states, such that $I\cdot a^{\ell_1} = I\cdot a^{\ell_3} \neq I \cdot a^{\ell_2}$, where $I$ is the set of initial states, and one of these sets is accepting and the other is not (otherwise they are equivalent). Note that the numbers can be guessed in binary. The matrix multiplication (fast exponentiation) can then be used to compute resulting sets of those transitions in polynomial time. Thus, we can check in \coNP whether the language of an NFA is $\R$-trivial.
    
    To show that $k$-$\R$-triviality is in \coNP, we first check in \coNP, given an NFA $\A$, whether the language $L(\A)$ is $\R$-trivial. If so, then it is $2^n$-$\R$-trivial by Theorem~\ref{thmMain}, since the depth of the minimal DFA is bounded by $2^n$, where $n$ is the number of states of $\A$. To show that $L(\A)$ is not $k$-$\R$-trivial, we need to find two $\sim^{\R}_{k}$-equivalent words such that exactly one of them belongs to $L(\A)$. Since every class defined by $a^\ell$, for $\ell < k$, is a singleton, we need to find $k < \ell \le 2^n$ such that $a^k \sim_k a^\ell$ and only one of them belongs to $L(\A)$. Since  $a^k \sim_k a^\ell$ holds for every $\ell > k$, this can be done in nondeterministic polynomial time by guessing $\ell$ in binary and using the matrix multiplication to compare the states reachable by $a^k$ and $a^\ell$ and verifying that one is accepting and the other is not.

    To show that both problems are \coNP-hard, we use the construction of~\cite{StockmeyerM73} that we recall here showing that universality is \coNP-hard for unary NFAs. Let $\varphi$ be a formula in 3CNF with $n$ distinct variables, and let $C_k$ be the set of literals in the $k$-th conjunct, $1 \le k \le m$. The assignment to the variables can be represented as a binary vector of length $n$. Let $p_1,p_2,\ldots,p_n$ be the first $n$ prime numbers. For a natural number $z$ congruent with 0 or 1 modulo $p_i$, for every $1\le i \le n$, we say that $z$ satisfies $\varphi$ if the assignment $(z \bmod p_1, z \bmod p_2,\ldots, z \bmod p_n)$ satisfies $\varphi$. Let 
    \[
      E_0 = \bigcup_{k=1}^{n} \bigcup_{j=2}^{p_k-1} 0^j\cdot (0^{p_k})^*
    \]
    that is, $L(E_0) = \{ 0^z \mid \exists k \le n, z \not\equiv 0 \bmod p_k \text{ and } z \not\equiv 1 \bmod p_k \}$ is the set of natural numbers that do not encode an assignment to the variables. For each conjunct $C_k$, we construct an expression $E_k$ such that if $0^z \in L(E_k)$ and $z$ is an assignment, then $z$ does not assign the value 1 to any literal in $C_k$. For example, if $C_k = \{x_{r}, \neg x_{s}, x_{t}\}$, for $1 \le  r,s,t \le n$ and $r,s,t$ distinct, let $z_k$ be the unique integer such that $0\le z_k < p_rp_sp_t$, $z_k \equiv 0 \bmod p_r$, $z_k \equiv 1 \bmod p_s$, and $z_k \equiv 0 \bmod p_t$. Then
    \[
      E_k = 0^{z_k} \cdot (0^{p_rp_sp_t})^*\,.
    \]
    Now, $\varphi$ is satisfiable if and only if there exists $z$ such that $z$ encodes an assignment to $\varphi$ and $0^z \notin L(E_k)$ for all $1\le k \le m$, which is if and only if $L(E_0 \cup \bigcup_{k=1}^{m} E_k) \neq 0^*$.
    This shows that universality is \coNP-hard for NFAs over a unary alphabet. Let $p_n^\# = \Pi_{i=1}^{n} p_i$. 
    If $z$ encodes an assignment of $\varphi$, then, for any natural number $c$, $z+c\cdot p_n^\#$ also encodes an assignment of $\varphi$. Indeed, if $z \equiv x_i \bmod p_i$, then $z + c\cdot p_n^\# \equiv x_i \bmod p_i$, for every $1\le i\le n$. This shows that if  $0^z \notin L(E_k)$ for all $k$, then $0^z (0^{p_n^\#})^* \cap L(E_0 \cup \bigcup_{k=1}^{m} E_k) = \emptyset$. Since both languages are infinite, the minimal DFA recognizing the language $L(E_0 \cup \bigcup_{k=1}^{m} E_k)$ must have a nontrivial cycle. Therefore, if the language is universal, then it is $k$-$\R$-trivial for any $k\ge 0$, and if it is non-universal, then it is not $\R$-trivial. This proves \coNP-hardness of $k$-$\R$-triviality for every $k\ge 0$.
  \qed\end{pf}

  We now briefly discuss the complexity of the problem if the language is given as a poNFA rather than an NFA. 

  \begin{theorem}\label{poNFAtoRtriv}
    Given a poNFA $\A$, the problems of deciding whether the language accepted by $\A$ is $\R$-trivial, or $k$-$\R$-trivial for a given $k\geq 0$, are both \PSpace-complete.
  \end{theorem}
  \begin{pf}
    The membership in \PSpace follows from Theorems~\ref{RtrivNFA} and~\ref{kRtrivNFA}. \PSpace-hardness can be shown by a slight modification of the proof of Theorem~\ref{thmMainPONFAS}. Let $M$ be a DTM and $x$ be an input. We construct a binary regular expression $R_x$ from $M$ and $x$ as in the proof of Theorem~\ref{thmMainPONFAS} with the modification as if $M$ had a self-loop in the accepting state $q_f$. That is, if $\#w_1\#\cdots\#w_m\#$ is the unique accepting computation of $M$ on $x$, we consider all words of the form $\#w_1\#\cdots\#w_m\#(w_m\#)^*$ as correct encodings of the accepting computation of $M$ on $x$. The binary regular expression $R_x$ and its corresponding binary poNFA $\A_x$ are then constructed as in the proof of Theorem~\ref{thmMainPONFAS}. If $M$ does not accept $x$, then $L(\A_x)=\{0,1\}^*$, which is a $k$-$\R$-trivial language for every $k\ge 0$. If $M$ accepts $x$, then $L(\A_x)=\{0,1\}^* \setminus \enc(\#w_1\#\cdots\#w_m\#(w_m\#)^*)$. Since $|\enc(w_m\#)| \ge 2$, the sequence of prefixes
    \[
      \left(\enc(\#w_1\#\cdots\#w_m\#(w_m\#)^i), \enc(\#w_1\#\cdots\#w_m\#(w_m\#)^i w_m)\right)_{i= 0}^{\infty}
    \]
    is infinite and alternates between non-accepted and accepted words of $\A_x$. Consequently, the minimal DFA equivalent to $\A_x$ must have a nontrivial cycle, which means that the language $L(\A_x)=\{0,1\}^* \setminus \enc(\#w_1\#\cdots\#w_m\#(w_m\#)^*)$ is not $\R$-trivial~\cite{BrzozowskiF80}. Therefore, the language of a binary poNFA $\A_x$ is ($k$-)$\R$-trivial if and only if $M$ does not accept $x$.
  \qed\end{pf}

  Notice that we have used a binary alphabet. For unary languages, we now show that the class of languages of unary poNFAs and unary $\R$-trivial languages coincide. 
  
  \begin{theorem}
    The classes of unary poNFA languages and unary $\R$-trivial languages coincide.
  \end{theorem}
  \begin{pf}
    Since every $\R$-trivial language is a poNFA language (see, for example, Theorem~\ref{rpoNFAsRlangs}), we only need to prove that unary poNFA languages are $\R$-trivial. Assume for the contrary that there is a poNFA language $L$ over the alphabet $\{a\}$ that is not $\R$-trivial. Then the minimal DFA for $L$ is not partially ordered, and hence it has a non-trivial cycle. In other words, there are $k\ge 0$ and $\ell\ge 2$ such that for every $m\ge 0$, $a^{k+m\ell}\in L$ and $a^{k+m\ell+1}\notin L$. However, if a unary poNFA accepts an infinite language, then there is an integer $n$ such that the poNFA accepts all words of length longer than $n$ (cf.~the proof of Theorem~\ref{ponfasUnary}). This contradicts the existence of $k$ and $\ell$.
  \qed\end{pf}

\section{Deciding Universality of rpoNFAs}\label{sec_rponfa_univ}

	In this section, we return to the universality problem for the case of rpoNFAs.
	We first show that we can indeed obtain the hoped-for reduction in complexity
	when using a fixed alphabet. For the general case, however, we can recover the
	same \PSpace lower bound as for poNFAs, albeit with a more involved proof.
	Even for fixed alphabets, we can get a \coNP lower bound:

  \begin{lemma}\label{lemma_coNPhard}
    The universality problem of rpoNFAs is \coNP-hard even when restricting to alphabets with two letters.
  \end{lemma}
  \begin{pf}
    The first part of the proof is adapted from~\cite{Hunt73}. We use a reduction from the complement of CNF satisfiability. Let $U=\{x_1,x_2,\ldots,x_n\}$ be a set of variables and $\varphi = \varphi_1 \land \varphi_2 \land \cdots \land \varphi_m$ be a formula in CNF, where every $\varphi_i$ is a disjunction of literals. Without loss of generality, we may assume that no clause $\varphi_i$ contains both $x$ and $\neg x$. Let $\neg \varphi$ be the negation of $\varphi$ obtained by the de Morgan's laws. Then $\neg\varphi = \neg\varphi_1 \lor \neg\varphi_2 \lor \cdots \lor \neg\varphi_m$ is in DNF. For every $i=1,\ldots,m$, define $\beta_i = \beta_{i,1}\beta_{i,2}\cdots\beta_{i,n}$, where 
    \[
      \beta_{i,j} = \left\{
        \begin{array}{ll}
          0+1 & \text{ if } x_j \text{ and } \neg x_j \text{ do not appear in } \neg\varphi_i\\
          0   & \text{ if } \neg x_j \text{ appears in } \neg\varphi_i\\
          1   & \text{ if } x_j \text{ appears in } \neg\varphi_i
        \end{array}
        \right.
    \]
    for $j=1,2,\ldots,n$. Let $\beta = \bigcup_{i=1}^{m} \beta_{i}$. Then $w\in L(\beta)$ if and only if $w$ satisfies some $\neg\varphi_i$. That is, $L(\beta) = \{0,1\}^n$ if and only if $\neg\varphi$ is a tautology, which is if and only if $\varphi$ is not satisfiable. Note that by the assumption, the length of every $\beta_{i}$ is exactly $n$.

  \begin{figure}
    \centering
    \begin{tikzpicture}[>=stealth',->,shorten >=1pt,node distance=2.1cm,
      every node/.style={fill=white,font=\small},double distance=1pt,
      state/.style={ellipse,minimum size=5mm,draw=black,initial text=}]
      \node[state,initial,accepting,circle] (0) {$0$};
      \node[state,accepting]  (3) [right of=0] {$\alpha_1$};
      \node[]                 (5) [right of=3] {$\cdots$};
      \node[state,accepting]  (6) [right of=5] {$\alpha_{n-1}$};
      \node[state]            (7) [right of=6] {$\alpha_{n}$};
      \node[state,accepting]  (8) [right of=7] {$\alpha_{n+1}$};
      \node[state,accepting]  (1) [above of=7,node distance=1.5cm] {$q_{i,n}$};
      \node[state,accepting]  (2) [below of=7,node distance=1.5cm] {$q_{j,n}$};
        \path
          (0) edge[snake it,out=30,in=180]  node {$\beta_i$} (1)
          (0) edge[snake it,out=330,in=180] node {$\beta_j$} (2)
          (0) edge node {$0,1$} (3)
          (3) edge node {$0,1$} (5)
          (5) edge node {$0,1$} (6)
          (6) edge node {$0,1$} (7)
          (7) edge node {$0,1$} (8)
          (8) edge[loop above] node[outer sep=1pt] {$0,1$} (8)
          ;
      \end{tikzpicture}
      \caption{The rpoNFA $\M$ from the proof of Lemma~\ref{lemma_coNPhard}}
      \label{rpoNFAn_lem20}
    \end{figure}
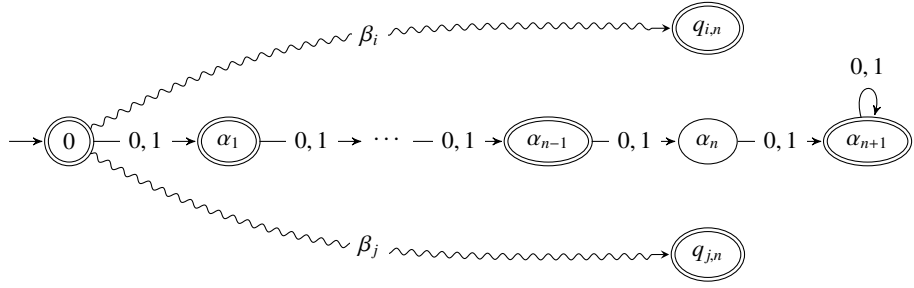

    We now construct an rpoNFA $\M$ as follows, see Figure~\ref{rpoNFAn_lem20}. The initial state of $\M$ is state $0$. For every $\beta_{i}$, we construct a deterministic path consisting of $n+1$ states $\{q_{i,0},q_{i,1},\ldots,q_{i,n}\}$ with transitions $q_{i,\ell+1} \in q_{i,\ell} \cdot \beta_{i,\ell}$ and $q_{i,0} = 0$ accepting the words $\beta_i$. In addition, we add $n+1$ states $\{\alpha_1,\alpha_2,\ldots,\alpha_{n+1}\}$ and transitions $\alpha_{\ell+1} \in \alpha_\ell \cdot a$, for $\ell < n+1$ and $\alpha_0 = 0$, and $\alpha_{n+1} \in \alpha_{n+1} \cdot a$, where $a\in \{0,1\}$, accepting all words of length different from $n$. The accepting states of $\M$ are the states $\{0,q_{1,n},\ldots,q_{m,n}\}\cup \{\alpha_1,\ldots \alpha_{n+1}\} \setminus \{\alpha_n\}$. Notice that $\M$ is restricted partially ordered. The automaton accepts the language $L(\M) = L(\beta) \cup \{w \in \{0,1\}^* \mid |w| \neq n\}$, which is universal if and only if $L(\beta) = \{0,1\}^n$.
  \qed\end{pf}

	For a matching upper bound, we use Lemmas~\ref{lemmaMinWord} and \ref{lemmaDLT}, which provide the main ingredients for
  showing that, if the size $|\Sigma|$ of the alphabet is bounded, then
  non-universality is witnessed by a word of polynomial length.
  Together with Lemma~\ref{lemma_coNPhard}, this allows us to establish the following result.

  \begin{theorem}\label{thmInclusionCoNP}
    Let $\Sigma$ be a fixed non-unary alphabet, and let $\B$ be an rpoNFA over $\Sigma$. If $\A$ is an NFA (poNFA, rpoNFA, DFA, poDFA) over $\Sigma$, then the problem whether $L(\A)\subseteq L(\B)$ is \coNP-complete.
  \end{theorem}
  \begin{pf}
    Hardness follows from Lemma~\ref{lemma_coNPhard} by letting $L(\A)=\Sigma^*$, which can be represented by a poDFA. 
    
    For membership, let $|\Sigma|=m$, and let $\A$ be an NFA. We show that $L(\A)$ is not a subset of $L(\B)$ if and only if there exists an NFA $\C$ of polynomial size with respect to $\B$ such that $L(\A) \cap L(\C) \neq \emptyset$ and $L(\B) \cap L(\C) = \emptyset$. Since such an NFA can be guessed by a nondeterministic algorithm, and the (non)emptiness of the intersection of the languages of two NFAs can be verified in polynomial time, we obtain that the problem whether $L(\A)\subseteq L(\B)$ is in \coNP.
    
    It remains to show that there exists such an NFA $\C$. Without loss of generality, we assume that $\B$ is complete; otherwise, we make it complete in polynomial time by adding a single sink state and the missing transitions. Let $k$ be the depth of $\B$. Then $k$ is bounded by the number of states of $\B$. By Theorem~\ref{thmMain}, language $L(\B)$ is $k$-$\R$-trivial, which means that it is a finite union of $\sim^{\R}_{k}$ classes. According to Lemmas~\ref{lemmaMinWord} and~\ref{lemmaDLT}, the length of the unique minimal representatives of the $\sim^{\R}_{k}$ classes is at most $\binom{k+m}{k}-1 < \frac{(k+m)^m}{m!}$. Since $m$ is a constant, the bound is polynomial in $k$. Now, $L(\A)$ is not a subset of $L(\B)$ if and only if there exists a word in $L(\A)$ that is not in $L(\B)$. This means that there exists a $\sim_k^{\R}$ class that intersects with $L(\A)$ and is disjoint from $L(\B)$. Let $w$ be the unique minimal representative of this class. In Theorem~\ref{kRtrivNFA}, we constructed a DFA $\D$ with at most $|w|+2$ states recognizing the class $w/_{\sim_k^{\R}}$. Notice that $\D$ is such that $L(\A) \cap L(\D) \neq \emptyset$, $L(\B) \cap L(\D) = \emptyset$, and that the size of $\D$ is polynomial with respect to the size of $\B$. This completes the proof.
  \qed\end{pf}

  \begin{corollary}\label{thmMainRPONFASfixed}
    Let $\Sigma$ be a fixed non-unary alphabet. Then the universality problem
    for rpoNFAs over $\Sigma$ is \coNP-complete.
  \end{corollary}
  \begin{pf}
    Hardness follows from Lemma~\ref{lemma_coNPhard}, the containment from Theorem~\ref{thmInclusionCoNP} by letting $L(\A)=\Sigma^*$.
  \qed\end{pf}

  Notice that the proof of Theorem~\ref{ponfasUnary} also applies to rpoNFAs, and hence we immediately have the following result.
  \begin{corollary}\label{rponfasUnary}
    The universality problem for rpoNFAs over a unary alphabet is \NL-complete.
    \qed
  \end{corollary}

	Without fixing the alphabet, universality remains \PSpace-hard even for rpoNFAs, but
	a proof along the lines of Theorem~\ref{thmMainPONFAS} is not straightforward.
	In essence, rpoNFAs lose the ability to navigate to an arbitrary position within a
	word for checking some pattern there. Expressions of the form $(\Sigma^*\cdots)$,
	which we frequently used, e.g., in \eqref{eq_re_wrong_enc}, are therefore excluded.
	This is problematic since the run of a polynomially space-bounded Turing machine may
	be of exponential length, and we need to match patterns across the full length of 
	our (equally exponential) encoding of this run. How can we navigate such a long
	word without using $\Sigma^*$?
	Our answer is to first define an rpoNFA that accepts all words except for a single,
	exponentially long word. This word will then be used as an rpoNFA-supported
	``substrate'' for our Turing machine encoding, which again follows Theorem~\ref{thmMainPONFAS}.
  
    \begin{table}
      \mbox{}\hfill%
        \begin{tabular}{l|lll}
          $k\backslash n$ & 1  & 2  & 3 \\\hline
          1 & $a_1$       & $a_1a_2$                          & $a_1a_2a_3$\\
          2 & $a_1^2$     & $a_1^2 a_2 a_1 a_2$               & $a_1^2 a_2 a_1 a_2 a_3 a_1a_2a_3$\\
          3 & $a_1^3$     & $a_1^3 a_2 a_1^2 a_2 a_1 a_2$           & $a_1^3 a_2 a_1^2 a_2 a_1 a_2 a_3 a_1^2 a_2 a_1 a_2 a_3 a_1a_2a_3$\\
          4 & $a_1^4$     & $a_1^4 a_2 a_1^3 a_2 a_1^2 a_2 a_1 a_2$ & $a_1^4 a_2 a_1^3 a_2 a_1^2 a_2 a_1 a_2 a_3 a_1^3 a_2 a_1^2 a_2 a_1 a_2 a_3 a_1^2 a_2 a_1 a_2 a_3 a_1a_2a_3$
        \end{tabular}%
      \hfill\mbox{}%
      \caption{Recursive construction of words $W_{k,n}$ as used in the proof of Lemma~\ref{exprponfas}}
      \label{tableWords}
    \end{table}

  \begin{lemma}\label{exprponfas}
    For all positive integers $k$ and $n$, there exists an rpoNFA $\A_{k,n}$ over an $n$-letter alphabet with $n(k+2)$ states such that the unique word not accepted by $\A_{k,n}$ is of length $\binom{k+n}{k}-1$.
  \end{lemma}
  \begin{pf}
    For integers $k,n\geq 1$, we recursively define words $W_{k,n}$ over the alphabet
    $\Sigma_n = \{a_1,a_2,\ldots, a_n\}$.
    For the base cases, we set $W_{k,1} = a_1^k$ and $W_{1,n} = a_1a_2\cdots a_n$.
    The cases for $k,n >1$ are defined recursively by setting
    \begin{align}
      W_{k,n} & = W_{k,n-1}\, a_{n}\, W_{k-1,n}\nonumber\\
				& = W_{k,n-1}\, a_n\, W_{k-1,n-1}\, a_n\, W_{k-2,n}\label{wkn}\\
				& = W_{k,n-1}\, a_n\, W_{k-1,n-1}\, a_n\, \cdots\, a_n\, W_{1,n-1}\, a_n\,. \nonumber
    \end{align}
    The recursive construction is illustrated in Table~\ref{tableWords}.
    The length of $W_{k,n}$ is $\binom{k+n}{n}-1$~\cite{dlt15}. Notice that $a_n$ appears exactly $k$ times in $W_{k,n}$.
    We further set $W_{k,n}=\eps$ whenever $kn=0$, since this is useful for defining $\A_{k,n}$ below.
    
    We construct an rpoNFA $\A_{k,n}$ over $\Sigma_n$ that accepts the language $\Sigma_n^*\setminus\{W_{k,n}\}$.
    For $n=1$ and $k\ge 0$, let $\A_{k,1}$ be the minimal DFA accepting the language $\{a_1\}^* \setminus \{a_1^k\}$.
    It consists of $k+2$ states of the form $(i;1)$ as depicted in Figure~\ref{rpoNFAn1}, together
    with the given transitions. 
    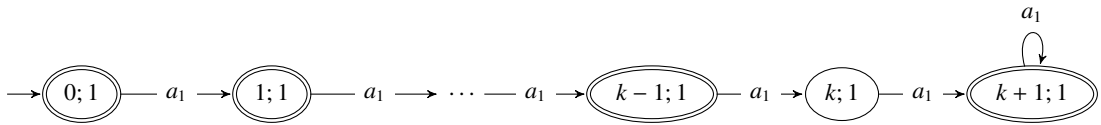
\begin{figure}[b]
      \centering
      \begin{tikzpicture}[>=stealth',bend angle=45,baseline,->,shorten >=1pt,node distance=2.5cm,
        every node/.style={fill=white,font=\small},double distance=1pt,
        state/.style={ellipse,minimum size=7mm,draw=black,initial text=}]
        \node[state,initial,accepting] (1) {$0;1$};
        \node[state,accepting]  (2)  [right of=1] {$1;1$};
        \node (3)  [right of=2] {$\ldots$};
        \node[state,accepting]  (k)  [right of=3] {$k-1;1$};
        \node[state]            (k1) [right of=k] {$k;1$};
        \node[state,accepting]  (k2) [right of=k1] {$k+1;1$};
          \path
            (1) edge node {$a_1$} (2)
            (2) edge node {$a_1$} (3)
            (3) edge node {$a_1$} (k)
            (k) edge node {$a_1$} (k1)
            (k1) edge node {$a_1$} (k2)
            (k2) edge[loop above] node[outer sep=1pt] {$a_1$} (k2) ;
      \end{tikzpicture}
      \caption{The rpoNFA $\A_{k,1}$ with $k+2$ states}
      \label{rpoNFAn1}
    \end{figure}
    All states but $(k;1)$ are final, and $(0;1)$ is initial. 
      
    Given $\A_{k,n-1}$, we recursively construct $\A_{k,n}$ as defined next.
    The construction for $n=2$ is illustrated in Figure~\ref{rpoNFAn2}.
  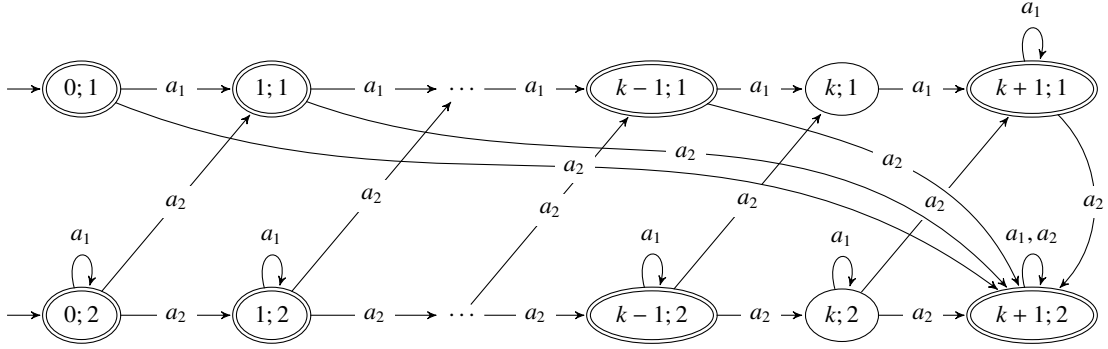
\begin{figure}
    \centering
    \begin{tikzpicture}[>=stealth',bend angle=45,baseline,->,shorten >=1pt,node distance=2.5cm,
      every node/.style={fill=white,font=\small},double distance=1pt,
      state/.style={ellipse,minimum size=5mm,draw=black,initial text=}]
      \node[state,initial,accepting] (0) {$0;1$};
      \node[state,accepting]  (1) [right of=0] {$1;1$};
      \node                   (2) [right of=1] {$\ldots$};
      \node[state,accepting]  (3) [right of=2] {$k-1;1$};
      \node[state]            (4) [right of=3] {$k;1$};
      \node[state,accepting]  (5) [right of=4] {$k+1;1$};
      \node[state,initial,accepting] (10) [below of=0,node distance=3cm] {$0;2$};
      \node[state,accepting]  (11) [right of=10] {$1;2$};
      \node                   (12) [right of=11] {$\ldots$};
      \node[state,accepting]  (13) [right of=12] {$k-1;2$};
      \node[state]            (14) [right of=13] {$k;2$};
      \node[state,accepting]  (15) [right of=14] {$k+1;2$};
        \path
          (0) edge node {$a_1$} (1)
          (1) edge node {$a_1$} (2)
          (2) edge node {$a_1$} (3)
          (3) edge node {$a_1$} (4)
          (4) edge node {$a_1$} (5)
          (5) edge[loop above] node[outer sep=1pt] {$a_1$} (5)
          (10) edge[loop above] node[outer sep=1pt] {$a_1$} (10)
          (11) edge[loop above] node[outer sep=1pt] {$a_1$} (11)
          (13) edge[loop above] node[outer sep=1pt] {$a_1$} (13)
          (14) edge[loop above] node[outer sep=1pt] {$a_1$} (14)
          (15) edge[loop above] node[outer sep=1pt] {$a_1,a_2$} (15)
          (13) edge node {$a_2$} (4)
          (5) edge[bend left] node {$a_2$} (15)
          (10) edge node {$a_2$} (11)
          (11) edge node {$a_2$} (12)
          (12) edge node {$a_2$} (13)
          (13) edge node {$a_2$} (14)
          (14) edge node {$a_2$} (15)
          (10) edge node {$a_2$} (1)
          (11) edge node {$a_2$} (2)
          (12) edge node {$a_2$} (3)
          (14) edge node {$a_2$} (5)
          (3) edge[out=345,in=114] node {$a_2$} (15)
          (0) edge[out=338,in=143] node {$a_2$} (15)
          (1) edge[out=340,in=132] node {$a_2$} (15)
          ;
      \end{tikzpicture}
      \caption{The rpoNFA $\A_{k,2}$ with $2(k+2)$ states}
      \label{rpoNFAn2}
    \end{figure}
    We obtain $\A_{k,n}$ from $\A_{k,n-1}$ by adding $k+2$ states
    $(0;n),(1;n),\ldots,(k+1;n)$, where $(0;n)$ is added to the initial states,
    and all states other than $(k;n)$ are added to the final states.
    $\A_{k,n}$ therefore has $n(k+2)$ states.
    
    The additional transitions of $\A_{k,n}$ consist of four groups:
    \begin{enumerate}
      \item\label{r1} Self-loops $(i;n)\xrightarrow{a_j}(i;n)$ for every $i=0,\ldots,k+1$
    and $a_j=a_1,\ldots,a_{n-1}$;
      \item\label{r2} Transitions $(i;n)\xrightarrow{a_n}(i+1;n)$ for every $i=0,\ldots,k$, and the self-loop $(k+1;n)\xrightarrow{a_n} (k+1;n)$;
      \item\label{r3} Transitions $(i;n)\xrightarrow{a_n}(i+1;m)$ for every $i=0,\ldots,k$ and $m=1,\ldots,n-1$;
      \item\label{r4} Transitions $(i;m)\xrightarrow{a_n}(k+1;n)$ for every accepting state $(i;m)$ of $\A_{k,n-1}$.
    \end{enumerate}
    
    The additional states of $\A_{k,n}$ and transitions (\ref{r1}) and (\ref{r2}) ensure acceptance of every word that does not contain exactly $k$ occurrences of $a_n$.
    The transitions (\ref{r3}) together with the transitions (\ref{r4}) ensure acceptance of all words in $(\Sigma_{n-1}^* a_n)^{i+1} L(\A_{k-(i+1),n-1})a_n \Sigma_n^*$ for which the word between the $(i+1)$-st and the $(i+2)$-nd occurrence of $a_n$ is not of the form $W_{k-(i+1),n-1}$, and hence not a correct subword of $W_{k,n} = W_{k,n-1} a_n \cdots a_n W_{k-(i+1),n-1} a_n \cdots\allowbreak a_n W_{1,n-1} a_n$.
    The transitions (\ref{r4}) ensure that all words with a prefix  $w\cdot a_n$ are accepted, where $w$ is any word  $\Sigma_{n-1}^*\setminus\{W_{k,n-1}\}$ accepted by $\A_{k,n-1}$.
    Together, these conditions ensure that $\A_{k,n}$ accepts every input other than $W_{k,n}$.
    
    It remains to show that $\A_{k,n}$ does not accept $W_{k,n}$, which we do by induction on $(k,n)$.
    We start with the base cases.
    For $(0,n)$ and any $n\geq 1$, the word $W_{0,n}=\eps$ is not accepted by $\A_{0,n}$,
    since the initial states $(0,m)=(k,m)$ of $\A_{0,n}$ are not accepting.
    Likewise, for $(k,1)$ and any $k\ge 0$, we find that $W_{k,1}=a_1^k$ is not accepted by $\A_{k,1}$ (Figure~\ref{rpoNFAn1}).

    For the inductive case $(k,n)\ge (1,2)$, assume $\A_{k',n'}$ does not accept $W_{k',n'}$ for any $(k',n') < (k,n)$; here $\le$ is the standard product order.
    We have $W_{k,n} = W_{k,n-1} a_n W_{k-1,n}$, and $W_{k,n-1}$ is not accepted by $\A_{k,n-1}$ by induction.
    In addition, there is no transition under $a_n$ from any non-accepting state of $\A_{k,n-1}$ in $\A_{k,n}$.
    Therefore, if $W_{k,n}$ is accepted by $\A_{k,n}$, it must be accepted in a run starting from the initial
    state $(0;n)$.
    Since $W_{k,n-1}$ does not contain $a_n$, we find that $\A_{k,n}$ can only reach the states
    $(0;n)\cdot W_{k,n-1} a_n =\{(1;m)\mid 1\le m \le n\}$ after reading $W_{k,n-1} a_n$. 
    These are the initial states of automaton $\A_{k-1,n}$, which does not accept $W_{k-1,n}$ by induction.
    Hence $W_{k,n}$ is not accepted by $\A_{k,n}$.
  \qed\end{pf}
    
  As a corollary, we find that there are rpoNFAs $\A=\A_{n,n}$ for which the shortest non-accepted word is exponential in the size of $\A$. Note that $\binom{2n}{n}\geq 2^n$.

  \begin{corollary}\label{expCor}
    For every integer $n\ge 1$, there is an rpoNFA $\A_{n}$ over an $n$-letter alphabet with $n(n+2)$ states such that the shortest word not accepted by $\A_{n}$ is of length $\binom{2n}{n}-1$. Therefore, any minimal DFA accepting the same language has at least $\binom{2n}{n}$ states.
  \end{corollary}
  \begin{pf}
    This is immediate from Lemma~\ref{exprponfas} by setting $n=k$. 
  \qed\end{pf}

  To simulate exponentially long runs of a Turing machine, we start from an encoding of runs
  using words $\#w_1\#\cdots\# w_m\#$ as in Theorem~\ref{thmMainPONFAS}, but we combine every
  letter of this encoding with one letter of the alphabet of $\A_{n}$.
  We then accept all words for which the projection to the alphabet of $\A_{n}$ is accepted by
  $\A_{n}$, i.e., all but those words of exponential length that are based on the unique
  word not accepted by $\A_{n}$. We ensure that, if there is an accepting run, it will have an
  encoding of this length. It remains to eliminate (accept) all words that correspond to a
  non-accepting or wrongly encoded run.
  We can check this as in Theorem~\ref{thmMainPONFAS}, restricting to the first components of
  our combined alphabet.
  The self-loop that was used to encode $\Sigma^*$ in poNFAs is replaced by a full copy of $\A_{n}$,
  with an additional transition from each state that allows us to leave this ``loop''.
  This does not simulate the full loop, but it allows us to navigate the entirety of our exponential
  word, which is all we need.

  \begin{theorem}\label{thmMainRPONFAS}
    The universality problem for rpoNFAs is \PSpace-complete.
  \end{theorem}
  \begin{pf}
    The membership follows since universality is in \PSpace for NFAs. 
    For hardness, we proceed as explained above.
    Consider a $p$-space-bounded DTM $M = \tuple{Q,T,\tmInputAlphabet,\gamma,\blank,q_o,q_f}$ as
    in the proof of Theorem~\ref{thmMainPONFAS}. We encode runs of $M$ as words
    over $T\times(Q\cup\{\eps\})\cup\{\#\}$ as before. We can use an unrestricted alphabet now, so 
    no binary encoding is needed, and the regular expressions can be simplified accordingly.
    
    If $M$ has an accepting run, then this run does not have a repeated configuration.
    For an input word $x$, there are
    $C(x) = (|T\times(Q\cup\{\eps\})|)^{p(|x|)}$ distinct configuration words in our encoding.
    Considering separator symbols $\#$, the maximal length of the encoding of a run without repeated configurations
    therefore is $1+ C(x)(p(|x|)+1)$, since every configuration word now ends with $\#$ and is thus of length $p(|x|)+1$.
    Let $n$ be the least number such that $|W_{n,n}|\geq 1+ C(x)(p(|x|)+1)$, where $W_{n,n}$ is the word from the proof of Lemma~\ref{exprponfas}.
    Since $|W_{n,n}|+1=\binom{2n}{n} \ge 2^n$, it
    follows that $n$ is smaller than $\left\lceil\log_2(1+ C(x)(p(|x|)+1))\right\rceil$ and hence polynomial
    in the size of $M$ and $x$.
    
    Consider the automaton $\A_{n,n}$ with alphabet $\Sigma_n=\{a_1,\ldots,a_n\}$ of
    Lemma~\ref{exprponfas}, and define $\Deltaplus = T\times(Q\cup\{\eps\})\cup\{\#,\$\}$.
    We consider the alphabet $\Pi=\Sigma_n\times\Deltaplus$, where the second letter is
    used for encoding a run as in Theorem~\ref{thmMainPONFAS}.
    Since $|W_{n,n}|$ may not be a multiple of $p(|x|)+1$, we add $\$$ to
    fill up any remaining space after the last configuration.
    For a word $w=\tuple{a_{i_1},\delta_1}\cdots \tuple{a_{i_\ell},\delta_\ell}\in\Pi^\ell$, we
    define $w[1]=a_{i_1}\cdots a_{i_\ell} \in \Sigma_n^\ell$ and $w[2]=\delta_1\cdots\delta_\ell\in\Deltaplus^\ell$.
    Conversely, for a word $v\in\Deltaplus^*$, we write $\enctwo(v)$ to denote the set of
    all words $w\in\Pi^{|v|}$ with $w[2]=v$. Similarly, for $v\in\Sigma_n^*$, $\enctwo(v)$
    denotes the words $w\in\Pi^{|v|}$ with $w[1]=v$. We extend this notation to sets of words.
     
    We say that a word $w$ encodes an accepting run of $M$ on $x$ if
    $w[1]=W_{n,n}$ and
    $w[2]$ is of the form $\#w_1\#\cdots\# w_m \# \$^j$ 
    such that there is
    an $i\in\{1,\ldots,m\}$ for which we have that
    
    \begin{itemize}
    \item $\#w_1\#\cdots\#w_i\#$ encodes an accepting run of $M$ on $x$ as in the proof of
    Theorem~\ref{thmMainPONFAS}, 
    \item $w_k=w_i$ for all $k\in\{i+1,\ldots,m\}$, and
    \item $j\leq p(|x|)$.
    \end{itemize}
    
    In other words, we extend the encoding by repeating the accepting configuration until we have
    less than $p(|x|)+1$ symbols before the end of $|W_{n,n}|$ and fill up the remaining places with $\$$.
    
    The modified encoding requires slightly modified expressions for capturing conditions (A)--(D) 
    from the proof of Theorem~\ref{thmMainPONFAS}. Condition (A) is not necessary, since we do not 
    encode symbols in binary.
    Condition (B) can use the same expression as in \eqref{eq_re_wrong_start}, adjusted to our
    alphabet:
    \begin{equation}
      \left(\varepsilon + \Pi + \Pi^2 +\cdots+ \Pi^{p(|x|)+1}\right) 
      + \sum_{0\leq j\leq p(|x|)+1} (\Pi^{j} \cdot \bar{E}_j\cdot\Pi^*)\label{eq_re_wrong_start_r}
    \end{equation}
    where $\bar{E}_j$ is the set $\Sigma_n\times(\Deltaplus\setminus\{v_j\})$ where $v_j$
    encodes the $j$-th symbol on the initial tape as in Theorem~\ref{thmMainPONFAS}.
    All uses of $\Pi^i$ in this expression encode words of polynomial length, which can be
    represented in rpoNFAs. Trailing expressions $\Pi^*$ do not lead to nondeterministic self-loops of Figure~\ref{fig_bad_pattern}.
    
    Condition (C) uses the same ideas as in Theorem~\ref{thmMainPONFAS}, especially the
    transition encoding function $f$, which we extend to $f: \Deltaplus^3\to \Deltaplus$.
    For allowing the last configuration to be repeated,
    we define $f$ as if the final state $q_f$ of $M$ had a self loop (a transition that does not modify
    the tape, state, or head position). Moreover, we generally permit $\$$ to occur instead of the
    expected next configuration symbol.
    We obtain:
    \begin{equation}
    \Pi^*\, \sum_{\delta_\ell,\delta,\delta_r \in \Deltaplus} \enctwo(\delta_\ell\delta\delta_r)\cdot  \Pi^{p(|x|)-1} \cdot \hat{f}(\delta_\ell,\delta,\delta_r)\cdot \Pi^*
    \label{eq_re_wrong_trans_r}
    \end{equation}
    where $\hat{f}(\delta_\ell,\delta,\delta_r)$ is $\Pi\setminus\enctwo(\{f(\delta_\ell,\delta,\delta_r),\$\})$.
    Expression \eqref{eq_re_wrong_trans_r} is not readily encoded in an rpoNFA, due to the leading $\Pi^*$.
    To address this, we replace $\Pi^*$ by the expression $\Pi^{\leq |W_{n,n}|-1}$, which matches
    every word $w\in\Pi^*$ with $|w|\leq |W_{n,n}|-1$. Clearly, this suffices for our case.
    As $|W_{n,n}|-1$ is exponential, we cannot encode this directly as for other expressions $\Pi^i$ before
    and we use $\A_{n,n}$ instead.
    
    In detail, let $E$ be the expression obtained from \eqref{eq_re_wrong_trans_r} when omitting the initial $\Pi^*$,
    and let $\A$ be an rpoNFA that accepts the language of $E$. We can construct $\A$ so that it has a single
    initial state. Moreover, let $\enctwo(\A_{n,n})$ be the automaton $\A_{n,n}$ of Lemma~\ref{exprponfas} with
    each transition $q\stackrel{a_i}{\to} q'$ replaced by all transitions
    $q\stackrel{\pi}{\to} q'$ with $\pi\in\enctwo(a_i)$.
    We construct an rpoNFA $\A'$ that accepts the language of $(\Pi^*\setminus\{\enctwo(W_{n,n})\}) + (\Pi^{\leq |W_{n,n}|-1}\cdot E)$
    by merging $\enctwo(\A_{n,n})$ with at most $n(n+2)$ copies of $\A$, where we identify the initial state of each
    such copy with a different final state of $\enctwo(\A_{n,n})$, if it does not introduce nondeterministic self-loops.
    The fact that $\enctwo(\A_{n,n})$ alone already accepts $(\Pi^*\setminus\{\enctwo(W_{n,n})\})$ was shown in
    the proof of Lemma~\ref{exprponfas}. This also implies that it accepts all words of length $\leq |W_{n,n}|-1$
    as needed to show that $(\Pi^{\leq |W_{n,n}|-1}\cdot E)$ is accepted.
    Entering states of (a copy of) $\A$ after accepting a word of length $\geq|W_{n,n}|$ is possible, but all words
    accepted in such a way are longer than $W_{n,n}$ and hence in $(\Pi^*\setminus\{\enctwo(W_{n,n})\})$.
  
    It remains to show that for every strict prefix $w_{n,n}$ of $W_{n,n}$, there is a state in $\A_{n,n}$ reached by $w_{n,n}$ that is the initial state of a copy of $\A$, and hence the check represented by $E$ in $\Pi^{\leq |W_{n,n}|-1}\cdot E$ can be performed. In other words, if $a_{n,n}$ denotes the letter following $w_{n,n}$ in $W_{n,n}$, then $w_{n,n}$ reaches a state in $\A_{n,n}$ that does not have a loop under $a_{n,n}$. However, this follows from the fact that $\A_{n,n}$ accepts everything but $W_{n,n}$, since then the DFA obtained from $\A_{n,n}$ by the standard subset construction has a path of length $\binom{2n}{n}-1$ labeled with $W_{n,n}$ without any loop. Moreover, any state of this path in the DFA is a subset of states of $\A_{n,n}$. Therefore, at least one of the states reachable under $w_{n,n}$ in $\A_{n,n}$ does not have a self-loop under $a_{n,n}$.
    
    Note that the acceptance of $(\Pi^*\setminus\{\enctwo(W_{n,n})\})$, which is a side effect of this encoding,
    does not relate to expressing \eqref{eq_re_wrong_trans_r} but is still useful for our intended overall encoding.
    
    The final condition (D) is minimally modified to allow for up to $p(|x|)$ trailing $\$$. 
    For a word $v$, we use $v^{\leq i}$ to abbreviate $(\varepsilon + v + \cdots + v^i)$,
    and we define $\bar{E}_f= (T\times (Q\setminus\{q_f\}))$ as before.
    Since not all words with too many trailing \$ are accepted by (C), we add this here instead.
    Moreover, we need to check that all the symbols $\$$ appear only at the end, that is, the last expression accepts all inputs where $\$$ is followed by a different symbol:
    \begin{align}
    &\Pi^* \enctwo(\#) (\Pi + \cdots + \Pi^{p(|x|)}) \enctwo(\$)^{\leq p(|x|)}  +{}\nonumber\\
    &\Pi^* \enctwo(\bar{E}_f) (\varepsilon + \Pi + \cdots + \Pi^{p(|x|)-1}) \enctwo(\#) \enctwo(\$)^{\leq p(|x|)} +{}\label{eq_re_wrong_final_r}\\
    &\Pi^* \enctwo(\$)^{p(|x|)+1} +{} \nonumber \\
    & (\Pi\setminus \enctwo(\$))^* \enctwo(\$) \enctwo(\$)^* (\Pi\setminus \enctwo(\$)) \Pi^* \nonumber
    \end{align}
    As before, we cannot encode the leading $\Pi^*$ directly as an rpoNFA, but we can perform
    a similar construction as in \eqref{eq_re_wrong_trans_r} to overcome this problem.
    
    The union of the rpoNFAs for \eqref{eq_re_wrong_start_r}--\eqref{eq_re_wrong_final_r} constitutes
    an rpoNFA that is polynomial in the size of $M$ and $x$, and that is universal if and only if $M$ does not accept $x$.
  \qed\end{pf}

\section{Inclusion and Equivalence of Partially Ordered NFAs}
\label{Sec:Inclusion}

  Universality is closely related to the inclusion and equivalence problems, which are of interest mainly from the point of view of optimization, e.g., in query answering. Given two languages $K$ and $L$ over $\Sigma$, the {\em inclusion problem\/} asks whether $K\subseteq L$ and the {\em equivalence problem\/} asks whether $K = L$. The relation of universality to inclusion and equivalence lies in the fact that the complexity of universality provides a lower bound on the complexity of both inclusion and equivalence. We now show that the complexities coincide, see Table~\ref{table_results_2}.

  The complexity of inclusion and equivalence for regular expressions of special forms has been investigated by Martens et al.~\cite{MartensNS09}. For a few of them, \PSpace-completeness of the inclusion problem has been achieved. The results are established for alphabets of unbounded size. Since some of the expressions define languages expressible by poNFAs, we readily have that the inclusion problem for poNFAs is \PSpace-complete. However, using Theorem~\ref{thmMainPONFAS} and the well-known \PSpace upper bound on inclusion and equivalence for NFAs, we obtain the following result.
  \begin{corollary}
    The inclusion and equivalence problems for poNFAs are \PSpace-complete even if the alphabet is binary.
  \end{corollary}
  
  The expressions in Martens et al.~\cite{MartensNS09} cannot be expressed as rpoNFAs. Hence the question for rpoNFAs was open. Using Theorem~\ref{thmMainRPONFAS} and the upper bound for NFAs, we can easily establish the following result.
  \begin{corollary}
    The inclusion and equivalence problems for rpoNFAs are \PSpace-complete.
  \end{corollary}
  
  If the alphabet is fixed, the complexity of inclusion (and of equivalence) is covered by Theorem~\ref{thmInclusionCoNP}.
  \begin{corollary}
    The inclusion and equivalence problems for rpoNFAs over a fixed alphabet are \coNP-complete.
  \end{corollary}
  
  Finally, for the unary case, it is known that the inclusion and equivalence problems for NFAs over a unary alphabet are \coNP-complete~\cite{HolzerK11,StockmeyerM73}. For poNFAs we obtain the following result.
  \begin{theorem}\label{inclusionpoNFA}
    The inclusion and equivalence problems for poNFAs over a unary alphabet are \NL-complete.
  \end{theorem}
  \begin{pf}
    The proof is a modification of the proof of Theorem~\ref{ponfasUnary}. Checking $L(\A)\subseteq L(\B)$ is easy if $L(\A)$ is finite, since there is at most $\depth(\A)+1$ strings to be checked. If $L(\A)$ is infinite, then there must be a simple path from an initial state to an accepting state via a state with a self-loop. Let $k$ denote the length of this path, which is bounded by the number of states. Then this path accepts all words of length at least $k$, that is, all words of the form $a^k a^*$. Then $L(\B)$ must also be infinite and, similarly, we get $\ell$ such that $a^\ell a^*$ all belong to $L(\B)$. For every $m\le \max\{k,\ell\}$, we check that if $a^m\in L(\A)$, then $a^m \in L(\B)$. This requires to perform $m+1$ nondeterministic logarithmic checks. As $m$ is smaller than the inputs, the proof is complete.
  \qed\end{pf}

\begin{table}
  \centering
  \begin{tabular}{@{}r@{\columnspace}r@{\columnspace}r@{\columnspace}r@{}}
      & {Unary alphabet} 
      & {Fixed alphabet} 
      & {Arbitrary alphabet}\\
    \hline
      DFA   & L-\complete       
            & \NL-\complete     
            & \NL-\complete      \\
    rpoNFA  & \NL-\complete     
            & \coNP-\complete    
            & \PSpace-\complete  \\
    poNFA   & \NL-\complete     
            & \PSpace-\complete  
            & \PSpace-\complete  \\
    NFA     & \coNP-\complete   
            & \PSpace-\complete 
            & \PSpace-\complete  
  \end{tabular}%
  \caption{Complexity of deciding inclusion and equivalence}\label{table_results_2}
\end{table}

\section{Deterministic Regular Expressions and Partially Ordered NFAs}

  In this section, we point out the relationship of partially ordered NFAs to deterministic regular expressions (DREs)~\cite{Bruggemann-KleinW98a}. DREs are of interest in schema languages for XML data -- Document Type Definition (DTD) and XML Schema Definition (XSD) -- since the World Wide Web Consortium standards require that the regular expressions in their specification must be deterministic.

  The {\em regular expressions\/} (REs) over an alphabet $\Sigma$ are defined as follows: $\emptyset$, $\eps$ and $a$, $a \in \Sigma$, are regular expressions. If $r$ and $s$ are regular expressions, then $(r\cdot s)$, $(r+s)$ and $(r)^*$ are regular expressions. The language defined by a regular expression $r$, denoted by $L(r)$, is inductively defined by $L(\emptyset)=\emptyset$, $L(\eps)=\{\eps\}$, $L(a)=\{a\}$, $L(r\cdot s)=L(r)\cdot L(s)$, $L(r+s) = L(r) \cup L(s)$, and $L(r^*) = \{\eps\}\cup \bigcup_{i=1}^{\infty} L(r)^i$, where $L(r) \cdot L(s)$ denotes the concatenation of the languages $L(r)$ and $L(s)$. Let $r$ be a regular expression, and let $\overline{r}$ be a regular expression obtained from $r$ by replacing the $i$-th occurrence of symbol $a$ in $r$ by $a_i$. For instance, if $r = (a+b)^* b (a+b)$, then $\overline{r} = (a_1+b_1)^* b_2 (a_2+b_3)$. A regular expression $r$ is {\em deterministic\/} (one-unambiguous~\cite{Bruggemann-KleinW98a} or DRE) if there are no words $wa_iv$ and $wa_jv'$ in $L(\overline{r})$ such that $i\neq j$. For instance, the expression $(a+b)^* b (a+b)$ is not deterministic since the strings $b_2a_2$ and $b_1 b_2a_2$ are both in $L((a_1+b_1)^* b_2 (a_2+b_3))$. A regular language is {\em DRE definable\/} if there exists a DRE that defines it. Br\"uggemann-Klein and Wood~\cite{Bruggemann-KleinW98a} showed that not every regular language is DRE definable.
  
  The important question is then whether a regular language is DRE definable. This problem has been shown to be \PSpace-complete~\cite{CzerwinskiDLM13}. Since the language of the expression $(a+b)^* b (a+b)$ is not DRE definable~\cite{Bruggemann-KleinW98a}, but it can be easily expressed by a poNFA, DRE definability is nontrivial for poNFAs. Its complexity however follows from existing results, namely from the proof in Bex et al.~\cite{BexGMN09} showing \PSpace-hardness of DRE-definability for regular expressions, since the regular expression constructed there can be expressed as a poNFA. Thus, we readily have the following:
  
  \begin{corollary}
    To decide whether the language of a poNFA is DRE definable is \PSpace-complete.
  \end{corollary}

  On the other hand, the problem is trivial for the languages of rpoNFAs, which makes rpoNFAs interesting for the XML schema languages.
  \begin{theorem}\label{Theorem-DRE-definability}
    Every rpoNFA language is DRE definable.
  \end{theorem}
  
  To prove the theorem, we need to introduce a few notions. For a state $q$ of an NFA $\A$, the {\em orbit of $q$\/} is the maximal strongly connected component of $\A$ containing $q$. State $q$ is called a {\em gate\/} of the orbit of $q$ if $q$ is accepting or has an outgoing transition that leaves the orbit. The orbit automaton of state $q$ is the sub-automaton of $\A$ consisting of the orbit of $q$ in which the initial state is $q$ and the accepting states are the gates of the orbit of $q$. We denote the orbit automaton of $q$ by $\A_q$. The orbit language of $q$ is $L(\A_q)$. The orbit languages of $\A$ are the orbit languages of states of $\A$.
  
  An NFA $\A$ has the {\em orbit property\/} if, for every pair of gates $q_1, q_2$ in the same orbit in $\A$, the following properties hold:
      (i) $q_1$ is accepting if and only if $q_2$ is accepting, and
      (ii) for all states $q$ outside the orbit of $q_1$ and $q_2$, there is a transition $q \in q_1 \cdot a$ if a and only if there is a transition $q\in q_2\cdot a$.

  Br\"uggemann-Klein and Wood~\cite{Bruggemann-KleinW98a} have shown that the language of a minimal DFA $\A$ is DRE-definable if and only if $\A$ has the orbit property and all orbit languages of $\A$ are DRE-definable.
    
  \begin{lemma}
    Every language of a minimal partially ordered DFA is DRE-definable.
  \end{lemma}
  \begin{pf}
    Every orbit of a partially ordered DFA is a singleton, and hence it satisfies the orbit property. The orbit language is either empty or $A^*$ for some alphabet $A$, and therefore DRE-definable.
  \qed\end{pf}
     
  \begin{pf}[of Theorem~\ref{Theorem-DRE-definability}]
    Every language defined by an rpoNFA is $\R$-trivial (Theorem~\ref{rpoNFAsRlangs}), and hence its minimal DFA is partially ordered~\cite{BrzozowskiF80}. By the previous lemma, the language is DRE-definable.
  \qed\end{pf}

  Finally, note that the converse of Theorem~\ref{Theorem-DRE-definability} does not hold. The expression $b^* a (b^* a)^*$ is deterministic~\cite{CzerwinskiDLM13} and it can be easily verified that its minimal DFA is not partially ordered. Therefore, the expression defines a language that is not $\R$-trivial.

\section{Conclusion}\label{conclusion}
  Our results regarding the complexity of deciding universality for partially ordered NFAs are summarized in Table~\ref{table_results}. We found that poNFAs over a fixed, two-letter alphabet are still powerful enough to recognize the language of all non-accepting computations of a \PSpace Turing machine. Restricting poNFAs further by forbidding nondeterministic self-loops, we could establish lower \coNP complexity bounds for universality for alphabets of bounded size. We can view this as the complexity of universality of rpoNFAs in terms of the size of the automaton when keeping the alphabet fixed. Unfortunately, the complexity is \PSpace-complete even for rpoNFAs over arbitrary (unbounded) alphabets. The proof uses an interesting construction where the encoding of a Turing machine computation is ``piggybacked'' on an exponentially long word, for which a dedicated rpoNFA is constructed.
  
  We have characterized the expressive power of rpoNFAs by relating them to the class of $\R$-trivial languages. It is worth noting that the complexity bounds we established for recognizing $\R$-triviality for a given NFA agrees with the complexity of the rpoNFA universality problem for both fixed and arbitrary alphabets. Our results on universality therefore extend beyond rpoNFAs to arbitrary NFAs that recognize $\R$-trivial languages.
  
  Moreover, the results on universality further extend to the complexity of inclusion and equivalence, and to the complexity of DRE-definability. Restricted poNFAs ($\R$-trivial languages) have been shown to be of interest in schema languages for XML data.
  
  Our work can be considered as a contribution to the wider field of studying subclasses of star-free regular languages. The Straubing-Th\'erien hierarchy provides a large field for interesting future work in this area.

\subsubsection*{Acknowledgements}
  The authors gratefully acknowledge very useful suggestions and comments of the anonymous referees.

\section*{\refname}
\bibliographystyle{elsarticle-harv}
\bibliography{mybib}

\end{document}